\documentclass[article,nofootinbib,superscriptaddress,prd]{revtex4-2}
\usepackage[colorlinks=true, pdfstartview=FitV, linkcolor=red, citecolor=blue,
urlcolor=blue]{hyperref}
\usepackage[dvipdfmx]{graphicx}
\usepackage{amsmath,amssymb,bm}
\usepackage{color}
\usepackage{slashed}

\renewcommand\a{\alpha}
\renewcommand\b{\beta}
\renewcommand\d{\delta}
\renewcommand\k{\kappa}
\renewcommand\l{\lambda}
\renewcommand\r{\rho}

\renewcommand\t{\tau}

\renewcommand\j{\psi}
\renewcommand\o{\omega}
\newcommand\e{\epsilon}
\newcommand\g{\gamma}

\newcommand\m{\mu}
\newcommand\n{\nu}
\newcommand\x{\xi}
\newcommand\p{\pi}

\newcommand\s{\sigma}
\newcommand\f{\phi}
\newcommand\w{\eta}

\renewcommand\L{\Lambda}

\renewcommand\S{\Sigma}
\renewcommand\O{\Omega}

\newcommand\F{\Phi}

\newcommand{\diag}{{\rm{diag}}}

\newcommand{\Tr}{{\rm Tr}}

\newcommand{\im}{{\rm{Im}}}
\newcommand{\bx}{{\mathbf x}}

\newcommand{\Lag}{\mathcal{L}}

\newcommand{\ud}{\mathrm{d}}
\newcommand{\ue}{\mathrm{e}}

\newcommand{\non}{\nonumber\\}

\newcommand\pt{\partial}

\newcommand{\eq}[1]{Eq.~(\ref{#1})}
\newcommand{\eqs}[2]{Eqs.~(\ref{#1})-(\ref{#2})}

\newcommand\lb{\left(}
\newcommand\rb{\right)}
\newcommand\ls{\left[}
\newcommand\rs{\right]}

\newcommand{\lan}{\langle}
\newcommand{\ran}{\rangle}

\newcommand\ra{\rightarrow}

\newcommand{\bv}{{\bf v}}

\newcommand{\jb}{{\bar \j}}

\renewcommand{\part}{{\rm part}}

\begin{abstract}
We study chiral symmetry breaking and restoration in accelerating and rotating frames using low-energy effective models. By analyzing the chiral condensate in Rindler coordinates, we show that different renormalization schemes lead to distinct conclusions in accelerating frame: the scheme with subtracting divergences in Rindler vacuum supports an acceleration-independent critical temperatures, while the other scheme with subtracting divergences in Minkowski vacuum suggests enhanced critical temperature. We further investigate system with both rotation and acceleration. We find that the critical acceleration (see definition in Section V) for chiral symmetry restoration decreases with angular velocity, indicating cooperative effects from acceleration-induced thermalization and rotation-induced effective chemical potential.

\end{abstract}

\begin{document}
\title{Chiral symmetry breaking in accelerating and rotating frames}
\author{Zhi-Bin Zhu}
\affiliation{Physics Department and Center for Particle Physics and Field Theory, Fudan University, Shanghai 200438, China}
\author{Hao-Lei Chen}
\email{hlchen@shu.edu.cn}
\affiliation{Department of Physics, Shanghai University, Shanghai 200444, China}
\affiliation{Key Laboratory of Nuclear Physics and Ion-beam Application (MOE), Fudan University, Shanghai 200433, China}
\affiliation{Shanghai Research Center for Theoretical Nuclear Physics, NSFC and Fudan University, Shanghai 200438, China}
\author{Xu-Guang Huang}
\email{huangxuguang@fudan.edu.cn}
\affiliation{Physics Department and Center for Particle Physics and Field Theory, Fudan University, Shanghai 200438, China}
\affiliation{Key Laboratory of Nuclear Physics and Ion-beam Application (MOE), Fudan University, Shanghai 200433, China}
\affiliation{Shanghai Research Center for Theoretical Nuclear Physics, NSFC and Fudan University, Shanghai 200438, China}

\maketitle

\section{Introduction}
Quantum Chromodynamics (QCD) is the fundamental theory of the strong interaction governing the dynamics of quarks and gluons. Extensive research on the QCD phase diagram aims to clarify the nature of strongly interacting quark-gluon matter under extreme conditions, including high temperature, high density, and strong external fields. In the low-temperature and low-baryon-density regime, quarks remain confined in hadrons (baryons and mesons), which corresponds to ordinary nuclear matter. Above a pseudo-critical temperature of approximately 150 MeV (at zero baryon density), quarks and gluons become deconfined and the (approximate) chiral symmetry is restored, resulting in the formation of a quark-gluon plasma (QGP) in which quarks and gluons can move over extended distances.
  
The QCD phase diagram is conventionally presented in the temperature ($T$)-baryon chemical potential ($\mu_B$) plane~\cite{Fukushima:2010bq}. Theoretical studies indicate a crossover transition at high temperatures with low baryon density, while a first-order phase transition emerges at high baryon densities with low temperatures. These regimes are connected by a critical endpoint (CEP) terminating the first-order phase boundary~\cite{Guenther:2020jwe}. The identification of this CEP in the $T$-$\mu_B$ plane remains a primary objective for both theoretical investigations and heavy-ion collision experiments~\cite{Bzdak:2019pkr}. Significant progress has been made in understanding the QCD phase structure through various theoretical approaches, including effective models~\cite{Buballa:2003qv,Fukushima:2008wg,Buballa:2014tba,Mannarelli:2019hgn}, lattice QCD simulations~\cite{Guenther:2020jwe,Ratti:2018ksb,Ding:2015ona,Aarts:2023vsf}, functional renormalization group~\cite{Pawlowski:2005xe,Drews:2016wpi,Dupuis:2020fhh,Fu:2019hdw,Fu:2022gou}, Schwinger–Dyson equation~\cite{Fischer:2018sdj,Gao:2016qkh,Guti_rrez_2014}, and holographic approaches~\cite{Rougemont:2023gfz,Zhao:2023gur,Liu_2024}. Recent theoretical developments have extended these studies to include additional background fields such as external electromagnetic fields~\cite{Andersen:2014xxa,Miransky:2015ava,Cao:2015cka} and global rotation~\cite{Chen:2021aiq}.

In recent years, the effects of rotation on QCD matter have attracted increasing attention, motivated by the presence of rapid rotation (or fluid vorticity) in non-central heavy-ion collisions~\cite{Huang:2020xyr,Huang:2020dtn} and neutron stars~\cite{Lattimer:2006xb,Paschalidis:2016vmz}. Following the early works~\cite{Chen:2015hfc,Jiang:2016wvv}, extensive efforts have been made to investigate the QCD phase structure under rotation~\cite{Chernodub:2016kxh,Chernodub:2017ref,Liu:2017spl,Huang:2017pqe,Zhang:2018ome,Wang:2018sur,Wang:2018zrn,Chen:2019tcp,Chen:2020ath,Chernodub:2020qah,Fujimoto:2021xix,Jiang:2021izj,Sadooghi:2021upd,Eto:2021gyy,Chen:2022mhf,Zhao:2022uxc,Chen:2022smf,Chen:2023cjt,Sun:2023kuu,Mameda:2023sst,Cao:2023olg,Eto:2023tuu,Eto:2023rzd,Sun:2024anu,Chen:2024tkr,Chen:2024jet,Wang:2024szr,Singha:2024tpo,Morales-Tejera:2025qvh,Kiefer:2025xdp,Nunes:2024hzy,Ambrus:2019ayb,Farias:2025vss}. Effective model studies suggest that rotation tends to restore chiral symmetry, whereas more recent lattice QCD simulations indicate that it enhances chiral symmetry breaking and confinement~\cite{Braguta:2021jgn,Chernodub:2022veq,Braguta:2023yjn,Yang:2023vsw,Braguta:2023iyx}. This apparent contradiction highlights the need for rigorous non-perturbative approaches to clarify the underlying mechanisms~\cite{Chen:2023cjt,Jiang:2024zsw,Wang:2025mmv,Fukushima:2025hmh}.

Compared to rotational effects, acceleration-related phenomena have received relatively less attention in QCD studies. According to the Unruh effect~\cite{hawking1975particle,unruh1976notes}, an observer undergoing constant proper acceleration $a$ perceives the Minkowski vacuum as a thermal bath with temperature~\cite{Crispino:2007eb} $$T_U = {a\over 2\pi}.$$ In Ref.~\cite{Kharzeev:2005iz}, the authors argued that partons in heavy-ion collisions experience characteristic deceleration on the order of the saturation scale, $a \sim Q_s \sim 1\text{GeV}$, leading to an Unruh temperature $T_U \sim 200\,\text{MeV}$ (see Ref.~\cite{Prokhorov:2025vak} for a numerical simulation of Unruh temperature in heavy-ion collisions). Since this exceeds the QCD pseudo-critical temperature, it suggests that acceleration-induced thermalization could have nontrivial consequences for the thermodynamics of the produced matter.

The general framework for quantum field theory in accelerating (Rindler) frame was established by Candelas et al.~\cite{candelas1976quantum,candelas1977vacuum,candelas1978fermion}, and Lee and collaborators \cite{Lee:1985rp} later demonstrating the universality of the Unruh effect for interacting quantum fields of arbitrary spin. These results provide a solid theoretical basis for studying acceleration-induced phenomena---such as chiral symmetry breaking and restoration---using effective models that incorporate thermodynamic behavior in non-inertial frames~\cite{Ohsaku:2004rv,Ebert:2006bh,Castorina:2012yg,Benic:2015qha,Casado-Turrion:2019gbg,Basu:2023bcu,Kou:2024dml,Chernodub:2025ovo}.
In addition, recent developments have employed Wigner function techniques~\cite{Becattini:2020qol,Palermo:2021hlf}, density operator formulations~\cite{Prokhorov:2018bql,Prokhorov:2019hif,Prokhorov:2019cik,Ambrus:2023smm}, and lattice simulation~\cite{Chernodub:2024wis} to investigate the thermodynamic properties of quantum systems in accelerating frame. These studies offer complementary insights into the behavior of QCD matter under acceleration.

Given that QGP created in heavy-ion collisions behaves as an ultra-vortical fluid with strong temperature gradients, in this paper we study the QCD matter subject to combined accelerating and rotational effects, with particular emphasis on acceleration-driven phenomena. We employ low-energy effective models in non-inertial frames to study the chiral condensate $\langle \bar{\psi}\psi\rangle$ (the order parameter for chiral symmetry breaking) as a function of acceleration $a$ and angular velocity $\omega$. Similar to rotational effects, the influence of acceleration on chiral symmetry is still under debate. Existing studies report conflicting conclusions: while some suggest that acceleration may restore chiral symmetry~\cite{Ohsaku:2004rv,Ebert:2006bh,Casado-Turrion:2019gbg}, others argue it may suppress it or have little effect~\cite{1984Acceleration,Salluce:2024jlj,Chernodub:2025ovo}. This discrepancy appears to stem primarily from the choice of renormalization schemes, as will be discussed in later sections.

The paper is organized as follows. Section~\ref{secare} discuss the general description of accelerating and rotating thermal equilibrium. Section~\ref{c2} establishes the Rindler coordinate framework for analyzing accelerating systems. Section~\ref{c4} and Sec.~\ref{c5} respectively investigate chiral symmetry breaking under pure acceleration and combined acceleration-rotation scenarios. Finally, Sec.~\ref{c6} concludes with a summary of the main results. We use natural units $\hbar=c=k_B=1$ and Minkowski metric $\eta^{\m\n}=\eta_{\m\n}=\diag(1,-1,-1,-1)$ through out this paper.

\section{Thermal equilibrium under acceleration and rotation}\label{secare}
Before we going into detailed discussion about how the acceleration and rotation affect the chiral symmetry breaking, it is important to notify that a many-body system could maintain thermal equilibrium under acceleration and rotation. We briefly discuss such a thermal equilibrium in this section; see Refs.~\cite{Becattini:2012tc,Becattini:2017ljh,Buzzegoli:2020ycf} for more discussions. The so-called local equilibrium density operator $\r_{\rm LE}$ is the density operator that maximizes the entropy functional
\begin{equation}\label{eq:entropy}
	S[\rho]=-\Tr\left(\rho\ln\rho\right),
\end{equation}
under the constraints that the local energy-momentum and angular momentum fluxes across a given space-like hypersurface $\Xi_\m$ in the Minkowski spacetime is fixed to their true physical values,
\begin{eqnarray}
	\label{eq:fixedt}
n_\m(x)\Tr\ls\rho T^{\m\n}(x)\rs=n_\m(x) T_{\rm true}^{\m\n}(x),\\
\label{eq:fixedj}
n_\m(x)\Tr\ls\rho J^{\m\r\s}(x)\rs=n_\m(x) J_{\rm true}^{\m\r\s}(x)
,
\end{eqnarray}
where $n_\m$ is the normal direction of $\Xi_\m$, $T^{\m\n}$ is the energy-momentum tensor, $J^{\m\r\s}=T^{\m\s} x^\r-T^{\m\r} x^\s+\Sigma^{\m\r\s}$ is the angular momentum tensor with $\S^{\m\r\s}$ the spin tensor. The resultant $\r_{\rm LE}$ has the following form:
\begin{equation}\label{eq:locequ}
	\rho_{\rm LE}=\frac{1}{Z_{\rm LE}} \exp\left[-\int d\Xi_\m \lb T^{\m\n}\b_\n-\frac{1}{2}J^{\m\r\s}\varpi_{\r\s}\rb\right],
\end{equation}
where $\b_\n$ and $\varpi_{\r\s} (=-\varpi_{\s\r})$ are two local Lagrange multipliers which have the physical meaning of thermal flow and ``chemical" potential for angular momentum and $Z_{\rm Le}$ is the partition function. When the system reaches global thermal equilibrium, the density operator must be time independent, or more precisely, independent of the choice of the space-like hypersurface $\Xi_\m$. This is achieved when $\b^\m$ and $\varpi^{\m\n}$ are constants as can be easily checked by noticing that $\partial_\m\lb T^{\m\n}\b_\n-J^{\m\r\s}\varpi_{\r\s}/2\rb=0$ owing to Gauss' theorem and the conservation laws for energy-momentum and angular momentum. Therefore, we have
\begin{equation}\label{eq:eqrho}
	\rho_{\rm eq}=\frac{1}{Z_{\rm eq}} \exp\left\{-\int d\Xi_\m \left[T^{\m\n}\lb \b_\n+\varpi_{\n\l}x^\l\rb-\frac{1}{2}\varpi_{\r\s}\Sigma^{\m\r\s}\right]\right\}.
\end{equation}

Let us choose $n_\m$ be along the direction of $\b_\m$ (which is time-like) and without loss of generality, we set $\b_\m$ to the coordinate time direction, $\b^\m=(\b, \bm 0)$, so that
\begin{eqnarray}\label{eq:eqrho2}
	\rho_{\rm eq}
		&=&\frac{1}{Z_{\rm eq}} \exp\left[-\b H  +\varpi_{0i} K^i+\frac{1}{2}\varpi_{ij}J^{ij}\right],
\end{eqnarray}
where $H=\int d^3\bx T^{00}$ is the Hamiltonian, $K^i=\int d^3 \bx \lb T^{0i} t-T^{00} x^i+\S^{00i}\rb$ is the boost operator, and $J^{ij}=\int d^3\bx\lb T^{0j}x^i- T^{0i}x^j+\Sigma^{0ij}\rb$ is the angular momentum operator. If we write $\varpi_{0i}=\b a^i$ and $\varpi_{ij}=\b\e^{ijk}\O^k$, it becomes clear that $\bm a$ is the linear acceleration and $\bm\O$ is the angular velocity. Note that though $K^i$ explicitly depends on $t$, it is actually time independent as it is conserved (which can be directly checked using Poincare algebra, $d K^i/dt=0$)~\cite{Becattini:2017ljh}. This allows us to set $t=0$ in $\r_{\rm eq}$. Choosing both the linear acceleration and rotation along $z$ direction, we have \begin{eqnarray}\label{eq:eqrho2}
	\rho_{\rm eq}
		&=&\frac{1}{Z_{\rm eq}} \exp\left[-\b \lb H  -a K^z-\O J^{z}\rb\right]\non
				&=&\frac{1}{Z_{\rm eq}} \exp\left[-\b \int d^3\bx \lb (1+az)T^{00} -a \S^{003}-\O(T^{02} x-T^{01}y+\S^{012})\rb\right]\bigg|_{t=0} ,
\end{eqnarray}
where $J^z=J^{12}$.

We now apply the above formalism to the single-flavor Nambu–Jona-Lasinio (NJL) model, whose Lagrangian is
\begin{equation}\label{eq:actddd}
	\mathcal{L}=\bar{\psi}\left[i \gamma^\mu\pt_\mu-m_0\right] \psi+\frac{G_\pi}{2}\left[(\bar{\psi} \psi)^2+\left(\bar{\psi} i \gamma^5 \psi\right)^2\right],
\end{equation}
where $\psi$ denotes the Dirac spinor field, $m_0$ is the fermion mass, and $G_\pi$ is the coupling constant. Applying the Noether theorem, we obtain the canonical energy-momentum tensor and spin tensor as
\begin{eqnarray}
	\label{eq:njltmn}
	T^{\m\n}&=&\jb i \g^\m \pt^\n\j-g^{\m\n}{\cal L},\\
	\label{eq:njlspindd}
	\S^{\m\r\s}&=&\frac{1}{2}\jb\g^\m\s^{\r\s}\j,
\end{eqnarray}
with $\s^{\m\n}=(i/2)[\g^\m, \g^\n]$ so that $\S^{003}=(i/2)\jb\g^3\j$ and $\S^{012}=(1/2)\j^\dag \s^z\j$. Then we have
\begin{eqnarray}
	\label{eq:njl:hd}
	H-a K^z-\O J^z&=&\int d^3 \bx\Big\{(1+az)\bar{\psi}\left[-i\gamma^{i} \partial_i+m_0 \right] \psi\non&&-(1+az)\frac{G_\p}{2} \left[(\bar{\psi} \psi)^2+\left(\bar{\psi} i \gamma_5 \psi\right)^2\right]-a\frac{i}{2}\jb\g^3\j-\O\j^\dag\ls -ix\pt_y+iy\pt_x+\frac{\s^z}{2}\rs\j\Big\}.
\end{eqnarray}
The partition function can then be written in a path-integral form as
\begin{eqnarray}\label{eq:njl:Z}
	Z_{\rm eq}=\Tr\; e^{-\b(H-a K^z-\O J^z)}=\int [d\psi][d\bar{\psi}]e^{-S_E[\psi,\bar{\psi}]},
\end{eqnarray}
where $S_E$ is the Euclidean action which is obtained from the Minkowski action $S=\int d^4x\jb i\g^0\pt_t\j-\int dt (H-a K^z-\O J^z)$ by the standard Wick rotation and imaginary-time compactification with anti-periodic boundary condition for $\j$ and $\jb$ in the imaginary-time direction. Notably, the Minkowski action $S$ is completely equal to the NJL action in an accelerating and rotating coordinates with metric
\begin{eqnarray}
\label{metricar}
	g_{\m\n}=
	\begin{pmatrix}
		(1+az)^2-\O^2 (x^2+y^2) & \O y & -\O x & 0 \\
		\O y & -1 & 0 & 0 \\
		-\O x & 0 & -1 & 0 \\
		0 & 0 & 0 & -1
	\end{pmatrix};
\end{eqnarray}
see Sec.~\ref{c5}, especially \eq{actioninar}. Therefore, we can study the accelerating and rotating thermal equilibrium by formulating the theory in the accelerating and rotating coordinates so that the techniques developed for field theory in curved spacetime can be used. This will be our strategy in the following sections and it will be shown more convenient to shift $z$ coordinate to $\x=z+1/a$ coordinate (see Sec.~\ref{c2}).  

It is well known that the definitions of energy-momentum tensor and spin tensor are not unique. They suffer from the so-called pseudo-gauge ambiguity~\cite{Becattini:2018duy,Huang:2024ffg}. For any differentiable local tensor $\F^{\l\m\n}(=-\F^{\l\n\m})$, $T'^{\m\n}$ and $\S'^{\m\r\s}$ defined through the following pseudo-gauge transformation,
\begin{eqnarray}
	\label{eq:psudog1}
	T'^{\m\n}&=& T^{\m\n}+\frac{1}{2}\pt_\l\lb \F^{\l\m\n}+\F^{\n\m\l}+\F^{\m\n\l}\rb,\\
    	\S'^{\m\r\s}&=&\S^{\m\r\s}-\F^{\m\r\s}, 
\end{eqnarray} 
are also well-defined energy-momentum and spin tensors in the sense that the conservation laws and the total four-momentum and total angular momentum are unchanged by this transformation. Therefore, the global thermal equilibrium density operator $\r_{\rm eq}$ and partition function $Z_{\rm eq}$ are invariant under pseudo-gauge transformation (though the local equilibrium density operator and partition function are not) and we thus have a freedom to choose a suitable pseudo-gauge to simply the action $S$. In particular, we could eliminate the acceleration-spin tensor coupling $-a\S^{003}=-i a\jb\g^3\j/2$ term in \eq{eq:njl:hd} by choosing the following pseudo-gauge potential $\F^{\l\m\n}$:   
\begin{equation}
\label{eqpseudog}
\Phi^{\l\mu\n}=\frac{i}{2}\bar{\psi}\left(\w^{\l\mu} \g^\n-\w^{\l\nu}\g^\m\right)\psi.
\end{equation}
This transforms \eqs{eq:njltmn}{eq:njlspindd} to
\begin{eqnarray}
	\label{eq:njltmnpr}
	T'^{\m\n}&=&\frac{i}{2}[\bar{\psi}\g^\mu\pt^\nu\psi-\bar\psi\overleftarrow{\pt}^\n\g^\mu\psi]-g^{\mu\nu}\mathcal{L}',\\
	\label{eq:njlspinddpr}
	\S'^{\m\r\s}&=&\frac{1}{4}\jb\{\g^\m,\s^{\r\s}\}\j=-\frac{1}{2}\e^{\m\r\s\n} \jb\g_\n\g_5\j,
\end{eqnarray}
where $\mathcal{L}'$ is a new Lagrangian,
\begin{equation}\label{eq:actdddaa}
	\mathcal{L}'=\jb\left[\frac{i}{2}(\g^\m \pt_\m-\overleftarrow{\pt}_\m\g^\m)-m_0\right] \psi+\frac{G_\pi}{2}\left[(\bar{\psi} \psi)^2+\left(\bar{\psi} i \gamma^5 \psi\right)^2\right].
\end{equation}
The difference is that $\S'^{003}=0$ so that $-a\S'^{003}$ term no longer appears in $H'-a K'^z-\O J'^z$:
\begin{eqnarray}
	\label{eq:njl:hddfda}
		H'-a K'^z-\O J'^z&=&\int d^3 \bx\Big\{(1+az)\bar{\psi}\left[\frac{1}{2}(-i\gamma^{i} \partial_i+i\g^i\overleftarrow{\pt}_i)+m_0 \right] \psi\non&&-(1+az)\frac{G_\p}{2} \left[(\bar{\psi} \psi)^2+\left(\bar{\psi} i \gamma_5 \psi\right)^2\right]-\O\j^\dag\ls -ix\pt_y+iy\pt_x+\frac{\s^z}{2}\rs\j\Big\}.
\end{eqnarray}
After integration by part, one finds that $H'-a K'^z-\O J'^z$ indeed equals to $H-a K^z-\O J^z$ in \eq{eq:njl:hd} (up to surface terms which are assumed to be vanished). Similar to discussions around \eqs{eq:njl:Z}{metricar}, the partition function of $H'-a K'^z-\O J'^z$ can be formulated as path integral with Lagrangian (\ref{eq:actdddaa}) in accelerating and rotating frames with metric (\ref{metricar}).


\section{Rindler coordinates}\label{c2}
The spacetime geometry in Rindler coordinates is characterized by the metric (see Appendix \ref{apprind} for a derivation)
\begin{equation}\label{eq.Rindler metric}
    \ud s^2=\xi^2 \ud t^2-\ud \xi^2-\ud x_1^2-\ud x_2^2.
\end{equation}
A particle at rest in Rindler coordinates corresponds to the one undergoing constant proper acceleration in Minkowski spacetime. For a particle under proper acceleration $a>0$, the worldline is given by $t = a\,\tau$, $\xi = 1/a$, where $\tau$ denotes the proper time. To describe finite-temperature systems, a Euclidean formulation can be obtained via analytic continuation $t_E = i t$. A system with proper acceleration $a$ and temperature $T$ is described by the Euclidean metric:
\begin{equation}\label{eq.eRindler metric}
    \ud s^2=-\ud s_E^2= -\frac{\xi^2}{\nu^2} \ud t^2_E-\ud \xi^2 -\ud x_1^2 -\ud x_2^2  .
\end{equation}
where $\nu = 2\pi T / a$. The Euclidean time $t_E$ satisfies $0<t_E<2\pi$. This geometry represents a two-dimensional conical manifold with angular deficit $2\pi(1 - \nu^{-1})$~\cite{Prokhorov:2023dfg}. To avoid a negative angular deficit, we impose $\nu \geq 1$ in this work, which leads to the condition $T \geq T_U$. When $T< T_U$, the angular deficit becomes negative and the geometric meaning of manifolds becomes unclear. Previous work has shown that negative energy manifests at temperatures $T < T_U$, indicating that it is an unstable state~\cite{Prokhorov:2019hif}. When $T = T_U$, the parameter $\nu = 1$ eliminates the angular deficit, resulting in a flat Euclidean manifold—an outcome consistent with the Unruh effect. A rigorous derivation can be found in Ref.~\cite{1984Acceleration}.
Let us clarify the roles of observer and observed systems. In Unruh's  scenario, an accelerating observer detects the Minkowski vacuum as a thermal bath at $ T = T_U $. 
In this paper, we consider such an accelerating observer, while the observed system is stationary in the oberver's frame with temperature $ T \geq T_U $.

Renormalization schemes are essential in this analysis, as different approaches may yield distinct physical outcomes \cite{Salluce:2024jlj}. We evaluate two choices of renormalization schemes corresponding two viewpoints about the subtracted vacuum.
One viewpoint suggests that, for accelerating observers, physical measurements should be referenced to their
perceived vacuum state (the state containing no detectable particles), which leads to the renormalization scheme
subtracting the Rindler vacuum expectation value $\langle \hat{O} \rangle_R$:
\begin{equation}\label{eq:re1}
    \langle \hat{O} \rangle_{ren} =\langle \hat{O} \rangle_\beta-\langle \hat{O} \rangle_R.
\end{equation}
Here $\langle \hat{O} \rangle_\beta$ denotes the thermal expectation value of an arbitrary operator $\hat{O}$ at temperature $T = 1/\beta$.
Note that the vacuum term ($\langle \hat{O} \rangle_R$) in this scheme is observer-dependent, varying with acceleration. An alternative viewpoint is that the vacuum should be frame-independent and measurements should be referenced to the Minkowski vacuum~\cite{Becattini:2017ljh}, leading to the second scheme, subtracting the Minkowski vacuum expectation value
$\langle \hat{O} \rangle_M$:
\begin{equation}\label{eq:re2}
  \langle \hat{O} \rangle_{ren} =\langle \hat{O} \rangle_\beta-\langle \hat{O} \rangle_M.
\end{equation}
The first renormalization scheme extends flat-space renormalization intuitively—accelerating observers naturally references their local vacuum rather than the Minkowski vacuum which they perceive as thermal. We therefore tentatively adopt it in subsequent sections. Nevertheless, the second scheme is compelling, preserving frame-independent observables and satisfying the Unruh-Weiss condition where thermal corrections vanish~\cite{1984Acceleration}. For comparison, we give the results based on this scheme in Sec.~\ref{ap0}.

\section{Chiral condensate under acceleration} \label{c4}
\subsection{NL$\sigma$M analysis}\label{secnlsm}
We first employ the nonlinear sigma model (NL$\sigma$M) to investigate the effect of acceleration on the chiral condensate. The NL$\sigma$M consists of $N$ pion fields $\pi^a$ ($a = 1, \dots, N$) and a single $\sigma$ field. The corresponding Lagrangian in a general curved spacetime takes the form:
\begin{equation}
    \mathcal{L}_{NL\sigma M}=-\frac{1}{2} \Phi^T \Box  \Phi-M^2_\pi f_\pi \sigma,
\end{equation}
where $f_\pi$ is the pion decay constant, and $\Phi^T = (\pi^a, \sigma)$ is a multiplet constrained by the nonlinear condition $\Phi^T \Phi = f_\pi^2$. The operator $\Box$ denotes the covariant d'Alembertian acting on scalar fields, defined as:
\begin{equation}
    \Box \phi \equiv g^{\mu\nu} \nabla_\mu \nabla_\nu \phi=\frac{1}{\sqrt{-g}} \partial_\mu \left[g^{\mu\nu}\sqrt{-g}\partial_\nu \phi\right],
\end{equation}
where $\nabla_\mu$ is the covariant derivative and $g \equiv \det(g_{\mu\nu})$.

To impose the nonlinear constraint within the path integral formalism, we insert a delta functional:
\begin{equation}
    Z=\int \mathcal{D}[\Phi] \delta[\Phi^T \Phi-f_\pi^2] \exp\left(i \int d^4x \mathcal{L}_{NL\sigma M}\right).
\end{equation}
The constraint is implemented by introducing an auxiliary field $\lambda(x)$. After integrating out the pion fields and taking the large-$N$ limit, we arrive at the following effective action:
\begin{equation}
    \Gamma[\sigma,\lambda]=\int \ud^4x \left( - \frac{1}{2} \sigma \Box  \sigma +\frac{\lambda}{2} (\sigma^2-f_\pi^2)+\frac{N}{2} \ln \frac{-\Box+\lambda}{-\Box}- M^2_\pi f_\pi \sigma \right).
\end{equation}
The gap equations are obtained by minimizing the effective action with respect to the fields $\sigma$ and $\lambda$:
\begin{equation}\label{eq:nlsigma-gap}
\begin{aligned}
        &\frac{\delta \Gamma}{\delta \sigma}= -\Box  \sigma+\lambda \sigma-M^2_\pi f_\pi=0, \\
       & \frac{\delta \Gamma}{\delta \lambda}= \frac{\sigma^2-f^2_\pi}{2}+\frac{N}{2} G(x,x;\lambda)=0,
\end{aligned}
\end{equation}
where the propagator $G(x,x';\lambda)$ satisfies 
\begin{equation}
   (-\Box +\lambda)_x G(x,x';\lambda)=\frac{1}{\sqrt{-g}}\delta^4(x-x'). 
\end{equation}
In this work, we consider the chiral limit $M_\pi = 0$, under which the gap equations naturally yield two distinct phases: the chiral restoration phase with $\sigma = 0$ and the spontaneously broken phase with $\sigma  \neq 0$. This framework establishes a direct correspondence between the chiral condensate $\langle \bar{\psi} \psi \rangle$ in a fermionic theory and the expectation value $\langle \sigma \rangle=\s$ solved from gap equations (\ref{eq:nlsigma-gap}).

To obtain the propagator, we first consider the solution of Klein-Gordon equation. For the Rindler metric given in Eq.~(\ref{eq.Rindler metric}), the Klein-Gordon equation takes the explicit form: 
\begin{equation}
\left(-\frac{1}{\xi^2}\frac{\partial^2}{\partial t^2}+\frac{\partial^2}{\partial \xi^2}+\frac{1}{\xi}\frac{\partial}{\partial \xi}+\frac{\partial^2}{\partial x_1^2}+\frac{\partial^2}{\partial x_2^2}-m^2\right)\phi=0,
\end{equation}
with $m$ the mass parameter. Solving this equation yields the eigenfunctions of d'Alembertian:
\begin{equation}
\phi=\frac{1}{\sqrt{2\omega}}\frac{1}{2\pi}\sqrt{\frac{2\omega\sinh\pi\omega}{\pi^2}}K_{i\o}(m_\perp\xi)\ue^{-i\omega t+i{\mathbf k}\cdot{\mathbf x}},
\end{equation}
where $\bx=(x_1,x_2)$ is the transverse coordinate, and we define the transverse mass squared as $m_\perp^2 = k^2 + m^2$ and $K_\nu(x)$ is the modified Bessel function of the second kind.
Using the integral identity for modified Bessel functions~\cite{candelas1978fermion}:
\begin{equation}
\begin{aligned}
& \int_0^{\infty} \ud  x K_\mu(\alpha x) K_\nu(\alpha x) x^l \\
& \quad=2^{l-2} \alpha^{-l-1} \Gamma(l+1)^{-1} \Gamma\left(\frac{l+\mu+\nu+1}{2}\right) \Gamma\left(\frac{l+\mu-\nu+1}{2}\right) \Gamma\left(\frac{l-\mu+\nu+1}{2}\right) \Gamma\left(\frac{l-\mu-\nu+1}{2}\right),
\end{aligned}
\end{equation}
(valid for $\mathrm{Re}(\alpha )>0$ and $\mathrm{Re}(l+\mu +\nu )>-1$), and following the Klein-Gordon inner product in Rindler coordinates ~\cite{candelas1976quantum,Crispino:2007eb}, we find that the eigenfunctions obey the orthonormality condition~\cite{candelas1976quantum}:
\begin{equation}
(\phi_i,\phi_j)_{\text{KG}}=2\o\int_0^\infty \ud\xi \int\ud^2{\mathbf x} \sqrt{-g}g^{00}\phi_{i}\phi_{j}=\d(i,j),
\end{equation}
where $i,j = (\omega, k_1, k_2)$ labels the eigenmodes of the field $\phi$.
Upon performing the secondary quantization, the scalar field $\phi$ becomes an operator:
\begin{equation}
\hat\phi=\int^\infty_0\ud \omega\int\frac{\ud^2{\mathbf k}}{2\pi}\sqrt{\frac{\sinh\pi\omega}{\pi^2}}K_{i\omega}(m_\perp\xi)(\hat a_{\omega,k}\ue^{-i\omega t+i{\mathbf k}\cdot{\mathbf x}}+\hat a_{\omega,k}^\dagger\ue^{i\omega t-i{\mathbf k}\cdot{\mathbf x}}),
\end{equation}
where $\hat a_{\omega,k}$ and $\hat a_{\omega,k}^\dagger$ are the annihilation and creation operators, respectively.  Particles and antiparticles in Rindler coordinates are defined with respect to the Killing vector $\partial/\partial t$ through their frequency $\omega$: modes with positive (negative) frequency correspond to particles (antiparticles).
It follows directly that the Euclidean propagator can be written as:
\begin{equation}\label{eq:bGe}
\begin{split}
G_E(x_E,x_E') &= \int^\infty_0 \mathrm{d}\omega \int\frac{\mathrm{d}^2\mathbf{k}}{(2\pi)^2}\frac{\sinh\pi\omega}{\pi^2}K_{i\omega}(m_\perp\xi)K_{i\omega}(m_\perp\xi')e^{-\omega|t_E-t_E'|+i\mathbf{k}\cdot(\mathbf{x}-\mathbf{x}')},
\end{split}
\end{equation}
where $t_E$ is the Euclidean Rindler time. We emphasize that the above expression for $G_E$ corresponds to the Rindler vacuum at zero temperature, serving as the baseline for the construction of the thermal propagator and the renormalization scheme discussed later.

In previous studies, the temperature has typically been fixed at $T=T_U$, a natural choice from the Unruh effect (see, e.g., Ref.~\cite{Casado-Turrion:2019gbg}). In this work, we extend the analysis by developing a generalized formalism valid for arbitrary temperatures satisfying {$T\geq T_U$}. 

It is important to note that Eq.~(\ref{eq:bGe}) describes the Euclidean propagator without boundary condition, corresponds to a system in the Rindler vacuum at zero temperature, from which we can construct a propagator at any temperature by imposing periodic boundary condition in Euclidean time.
Specifically, the finite-temperature propagator $G_\nu$  should satisfy the condition $G_\nu(t_E,t_E')=G_\nu(t_E+\beta_R,t_E')$ where $\beta_R=1/T_R=a/T$ and $\nu=2\pi T/a$ as defined in Eq.~(\ref{eq.eRindler metric}).

Based on the periodic boundary condition in Euclidean time, the finite-temperature propagator $G_\nu$ can be constructed from the zero-temperature propagator $G_E$ in Eq.~(\ref{eq:bGe}) via
\begin{equation}\label{eq.scalarg}
\begin{split}
G_{\nu}(x_E, x'_E)&=\sum_n G_E(t_E-t_E'+\beta_R n)\\
&=\int^\infty_0\ud \omega\int\frac{\ud^2{\mathbf k}}{(2\pi)^2}\frac{\cosh\omega(|t_E-t_E'|-\beta_R /2)}{\sinh (\beta_R \omega/2)}\frac{\sinh \pi \omega}{\pi^2}K_{i\omega}(m_\perp\xi)K_{i\omega}(m_\perp\xi')\ue^{i{\mathbf k}\cdot({\mathbf x}-{\mathbf x'})}.\\
\end{split}
\end{equation}
Alternatively, this result can be obtained by directly solving the thermal Green's function equation in Euclidean Rindler space with periodic boundary condition in $t_E$:
\begin{equation}\label{eq:sdg}
\left(\frac{1}{\xi^2} \frac{\partial^2}{\partial t_E^2} + \frac{\partial^2}{\partial \xi^2} + \frac{1}{\xi} \frac{\partial}{\partial \xi} + \frac{\partial^2}{\partial x_{E1}^2}+\frac{\partial^2}{\partial x_{E2}^2} - m^2 \right) G(x_E,x'_E) = -\frac{\delta^{(4)}(x_E - x'_E)}{\xi},
\end{equation}
whose solution is
\begin{equation}\label{eq.scalarg2}
    G_\nu(x_E,x'_E)=\sum_n\int^\infty_0\frac{\ud \omega}{\beta}\int\frac{\ud^2{\mathbf k}}{(2\pi)^2}\frac{\ue^{i\omega_n(t_E-t_E')+i{\mathbf k}\cdot({\mathbf x}_1-{\mathbf x}_2)}}{\omega_n^2+\omega^2}\frac{2\omega\sinh\pi\omega}{\pi^2}K_{i\omega}(m_\perp\xi)K_{i\omega}(m_\perp\xi'),
\end{equation}
where $\omega_n=2\pi n/\beta_R$ is the Matsubara frequency of bosons. Performing the frequency summation in Eq.~(\ref{eq.scalarg2}) yields Eq.~(\ref{eq.scalarg}). 

For our purpose of studying chiral phase transition, the key quantity is the diagonal part of the propagator:
\begin{equation}\label{eq.botrg}
\begin{split}
        G_{\nu}(x_E,x_E)&=\int^\infty_0\frac{\ud \omega}{\pi^2}\frac{\ud^2{\mathbf k}}{(2\pi)^2} \coth{(\omega \beta_R/2)}\,\sinh(\pi\omega)  K^2_{i\omega}(m_\perp\xi).
\end{split}
\end{equation}
According to the first gap equation in Eq.~(\ref{eq:nlsigma-gap}), the order parameter $\sigma$ satisfies
$-\Box \sigma + \lambda \sigma = 0 $
in the chiral limit ($M_\pi \to 0$). Although both $\lambda$ and $\sigma$ generally depend on position due to the inhomogeneous nature of acceleration, we focus on phase transitions evaluated at the observer’s instantaneous location (with a fixed $\xi$). Under the slow-variation approximation ($\Box\sigma \approx 0$), the gap equation simplifies to $\l=0$ and $\s^2=f_\p^2-NG_\n(x_E,x_E; m=0)$, and the phase structure can be determined by evaluating the diagonal part of the massless scalar propagator at the observer's location. Applying the integral identity,
\begin{equation}
\begin{aligned}
& \int_0^{\infty} \ud  x K_{i\mu}(\alpha x) K_{i\mu}(\alpha x) x=\frac{\pi  \mu  }{2 \alpha ^2\sinh(\pi  \mu )},
\end{aligned}
\end{equation}
valid for $-1<\mathrm{Im}(\mu )<1$ and $\mathrm{Re}(\alpha )>0$, we can derive the diagonal part of the massless propagator as:
\begin{equation}\label{eq.s_trg_method1}
\begin{aligned}
     G_{\nu}(x_E,x_E;m=0)&=\int^\infty_0\frac{\ud \omega}{\pi^2}\frac{1}{(2\pi)} \coth{(\omega \beta_R/2)} \frac{\pi \omega}{2 \xi^2} \\
     &=\int^\Lambda_0{\ud \omega}\frac{\omega}{4 \pi^2 a^2 \xi^2}+\int^\infty_0{\ud \o} \frac{\omega}{4 \pi^2 a^2\xi^2}\frac{2}{e^{\omega\beta}-1}. 
\end{aligned}
\end{equation}
In the last line, we have applied a change of variables $\omega \to \omega/a$, which facilitates ultraviolet regularization via an $a$ independent energy cutoff $\Lambda$. 
To obtain a finite result, we subtract the divergent, temperature-independent part of $G_\nu$ (the first term in the second line of  \eq{eq.s_trg_method1}), corresponding to the vacuum contribution. This accounts to the a subtraction $G_\n^{ren}=G_\n-G_{\n=0}$. A more detailed discussion of the renormalization scheme can be found in Sec.~\ref{ap0}.
Accordingly, the renormalized diagonal part of the propagator takes the form:
\begin{equation}
    G^{ren}_{\nu}(x_E,x_E;m=0)=\frac{\nu^2}{48\pi^2\xi^2}.
\end{equation}
Note that this subtraction procedure can also be viewed as a renormalization procedure to the phenomenological constant, $f^2_\pi\to f^2_\pi+\frac{N\Lambda^2}{8\pi^2 a^2 \xi^2}$. Substituting this into the gap equation~\eqref{eq:nlsigma-gap}, we obtain the expression for the order parameter $\sigma$:
\begin{equation}\label{eq:sclar sigma}
\begin{split}
    \sigma^2=f_\pi^2-\frac{T^2}{a^2\xi^2}\frac{N}{12}.
\end{split}
\end{equation}
This shows that, as the temperature increases, the chiral condensate $\sigma$ decreases and eventually vanishes.
We define the critical temperature  at which $\sigma$ reaches $0$ as
\begin{equation}\label{eq:Tcboson}
\begin{split}
    T_c(\xi)=a\xi\sqrt{\frac{12f_\pi^2}{N}}.
\end{split}
\end{equation}
This result is consistent with earlier study~\cite{Casado-Turrion:2019gbg}.

As emphasized previously, our analysis focuses on the observer’s instantaneous local position. The worldline of an observer of constant proper acceleration $a$ is given by $t = a\tau$ and $\xi = 1/a$, as discussed in Sec.~\ref{c2}.
Thus we find that the critical temperature observed by this observer becomes independent of its proper acceleration: $T_c(\xi=1/a)=\sqrt{12f_\pi^2/N}\equiv T_{c0}$, indicating that the phase transition depends solely on the local temperature experienced by the accelerated observer, not on the proper acceleration of the observer itself. Note that thus \eq{eq:Tcboson} can be rewritten as $T_c(\x)/a\xi= T_{c0}$ with the left-hand side representing the Tolman-Ehrenfest critical temperature. Therefore the chiral phase transition happens at the same Tolman-Ehrenfest temperature showing that local observers at different $\xi$ agree with the critical temperature for chiral phase transition. We will revisit and further interpret this result after evaluating the fermionic case in the next subsection.

\subsection{NJL model analysis}\label{subsc.njl}
We next turn to the quark degree of freedom to investigate how acceleration affects fermions.
In contrast to the scalar case, a key distinction lies in the coupling between acceleration and fermion spin for some choices of pseudo-gauge, analogous to the spin-rotation coupling in a rotating system. 
Here we start with the single-flavor NJL model, whose dynamics in a general spacetime are governed by the curved-spacetime version of Lagrangian (\ref{eq:actddd}):
\begin{equation}\label{eq.NJLL}
\mathcal{L}_{N J L}=\bar{\psi}\left[i \gamma^\mu\nabla_\mu-m_0\right] \psi+\frac{G_\pi}{2}\left[(\bar{\psi} \psi)^2+\left(\bar{\psi} i \gamma^5 \psi\right)^2\right].
\end{equation}
The covariant derivative $\nabla_\mu$, which encodes the effects of curved spacetime on fermions, will be defined explicitly in the following discussion.
From the Lagrangian in Eq.~(\ref{eq.NJLL}), the corresponding generating functional or partition function is given by:
\begin{equation}\label{Eq,partion}
\begin{aligned}
Z & =\int \mathcal{D}[\bar{\psi}, \psi] \exp \left(i \int \ud^4 x \sqrt{-g} \mathcal{L}_{\mathrm{NJL}}\right) \\
& =\int \mathcal{D}[\bar{\psi}, \psi, \sigma, \pi] \exp \left\{i \int \ud^4 x \sqrt{-g}\left[\bar{\psi}\left(i \gamma^\mu \nabla_\mu-m-i \gamma^5 \pi \right) \psi-\frac{\sigma^2+\pi^2}{2 G_\pi}\right]\right\},
\end{aligned}
\end{equation}
where the Hubbard-Stratonovich transformation introduces auxiliary fields $\sigma \equiv -G_\pi\bar{\psi}\psi$ and $\pi \equiv -G_\pi\bar{\psi}i\gamma^5\psi$, with the constituent quark mass defined as $m = m_0 + \sigma$.
Within the mean-field approximation, we derive the spacetime-dependent effective potential density, which is given by:
\begin{equation}
        V_{eff}(x)=\frac{\sigma^2+\pi^2}{2 G_\pi} -\frac{1}{i} \operatorname{Tr}  \ln \left(i \gamma^\nu \nabla_\nu-m-i \gamma^5 \pi\right).
\end{equation}
Minimizing the effective potential density leads to the gap equations:
\begin{equation}\label{eq:gap simple}
        \frac{m-m_0}{G_\pi}=i\operatorname{Tr}(S), \quad  \frac{\pi}{G_\pi}=i\operatorname{Tr}(i\gamma^5 S),
\end{equation}
where the fermion propagator $S$ satisfies:
\begin{equation}\label{eq:EqSF}
    \left(i \gamma^\nu \nabla_\nu-m-i \gamma^5 \pi \right)_x S(x,x')=\frac{1}{\sqrt{-g}}\delta^4(x,x').
\end{equation}
Without loss of generality, here we can set $\pi=0$ due to the $U(1)$ chiral symmetry and focus solely on $\s$. 
Following general formalism of quantum field theory in curved spacetime, $\gamma^\mu=e^\mu_{\hat{m}}\gamma^{\hat{m}}$ represent the spacetime-dependent gamma matrices and $\gamma^{\hat{m}}$ represent the gamma matrices in Minkowski spacetime. And we choose the vierbein $e^\mu_{\hat{m}}$ as
\begin{equation}
     e^0_{\hat{0}}= \frac{1}{\xi}\,,\,e^i_{\hat{0}}=0,e^0_{\hat{j}}=0, e^i_{\hat{j}}=\delta^i_j.
\end{equation}
where $i,j=1,2,3$ with $3$ for $\xi$ coordinate. The covariant derivative $\nabla_\mu$ and spin connection $\Gamma_\mu$ satisfy:
\begin{equation}
    \begin{aligned}
        \nabla_\mu = \partial_\mu + \Gamma_\mu, \quad \Gamma_\mu = -\frac{i}{4} \omega_{\mu \hat{\n}\hat{\r}} \sigma^{\hat{\n} \hat{\r}}, \quad \sigma^{\m\n} = \frac{i}{2}[\gamma^{\hat{\m}},\gamma^{\hat{\n}}], \quad
        \omega_{\mu \hat{\k} \hat{\w}} = g_{\nu \rho}e^\nu_{\hat{\k}} \nabla_\mu e^\rho_{\hat{\eta}}.
    \end{aligned}
\end{equation}
For the metric specified in Eq.~(\ref{eq.Rindler metric}), the corresponding gamma matrices take the form:
\begin{equation}
\begin{gathered}
\gamma^0(x) = \frac{1}{\xi} {\gamma^{\hat{0}}}, \quad
\gamma^i(x) = \gamma^{\hat{i}},
\end{gathered}
\end{equation}
with the non-zero spin connection component:
\begin{equation}
    \Gamma_0 = \frac{1}{2} \gamma^{\hat{0}} \gamma^{\hat{3}}.
\end{equation}

Note that the spin connection introduces a spin-dependent term in the partition function [Eq.~(\ref{Eq,partion})] and in the propagator [see Eq.~(\ref{eq:EqSF}) and also Eq.~(\ref{eqgreenfuns}) below]. Such a term does not appear in \cite{Becattini:2017ljh,Prokhorov:2019cik}, where accelerating fermions are treated via the density operator. In fact, this extra contribution can be removed through a pseudo-gauge transformation, as we have shown in Sec.~\ref{secare}. The resultant theory is equivalent to replacing $\Lag_{NJL}$ in \eq{eq.NJLL} by $\Lag'_{NJL}=(\Lag_{NJL}+\Lag_{NJL}^\dag)/2$. Note that the equation of motion is unchanged by the pseudo-gauge transformation, and therefore both $\Lag'_{NJL}$ and $\Lag_{NJL}$ yield the same propagator. We will hance focus solely on $\Lag_{NJL}$ in the following.

We now return to the discussion of the propagator. Based on the above setup, the explicit expression of Eq.~(\ref{eq:EqSF}) is:
\begin{equation}
	\label{eqgreenfuns}
    {\left[\frac{\gamma^{\hat{0}}}{\xi}\left(i \partial_0+\frac{i}{2} \gamma^{\hat{0}} \gamma^{\hat{3}}\right)+i \gamma^{\hat{i}} \partial_i-m\right]S(x,x')\equiv (\hat D-m)S(x,x')=\frac{1}{\sqrt{-g}}\delta^4(x,x').}
\end{equation}
Since both $\Lag'_{NJL}$ and $\Lag_{NJL}$ yield the same equations of motion, $S(x,x')$ is also the propagator for $\Lag'_{NJL}$.
To solve this equation, we adopt the ansatz that the propagator can be expressed as
\begin{equation}
    S(x,x')=(\hat D+m)G(x,x'),
\end{equation}
with $G(x,x')$ an auxiliary propagator which satisfies
\begin{equation}\label{eq.defG}
\begin{aligned}
& \left(\hat{D}^2-m^2\right) G(x, x')=\frac{1}{\sqrt{-g}} \delta(x, x'),
\end{aligned}
\end{equation}
where the squared Dirac operator $\hat{D}^2$ explicitly takes the form:
\begin{equation}\label{eq.d2}
    \hat{D}^2= -\frac{1}{\xi^2}\left(i(i\partial_0)- \frac{\gamma^{\hat{0}} \gamma^{\hat{3}}}{2} \right)^2+\partial_3^2+\frac{1}{\xi}  \partial_3+\partial_1^2+\partial_2^2.
 \end{equation}
In comparison to the scalar case, a only distinguishing feature is the presence of the $\gamma^{\hat{0}}\gamma^{\hat{3}}$ term, which reflects the contribution from spin. Therefore, the propagator can be constructed by the eigenmodes of Klein-Gordon operator with $i\o\ra i \omega - {\gamma^{\hat{0}} \gamma^{\hat{3}}}/{2}$:
\begin{equation}
    e^{-i \omega t}e^{i \bold{k} \cdot \bold{x}} K_{i \omega - {\gamma^{\hat{0}} \gamma^{\hat{3}}}/{2}} (m_\perp \xi).
\end{equation}
For later use, it is useful to notice that, for a function containing $\gamma^0 \gamma^3$, it can always be decomposed as
 \begin{equation}
     f(\gamma^0 \gamma^3)= P^+ f(1) + P^- f(-1),
 \end{equation}
 where $P^{\pm}=(1\pm \gamma^0 \gamma^3)/2$ are two projections. 
Therefore, the auxiliary propagator in Euclidean Rindler spacetime is directly obtained from \eq{eq:bGe}. The result is 
\begin{equation}
     G_E(x_E,x_E')=\int^\infty_0 \mathrm{d}\omega\int\frac{\mathrm{d}^2{\mathbf k}}{(2\pi)^2}\frac{\sinh\pi(\omega+i \gamma^{\hat{0}} \gamma^{\hat{3}}/2)}{\pi^2}K_{i\omega- \gamma^{\hat{0}} \gamma^{\hat{3}}/2}(m_\perp\xi)K_{i\omega- \gamma^{\hat{0}} \gamma^{\hat{3}}/2}(m_\perp\xi')e^{-\omega|t_E-t_E'|+i{\mathbf k}\cdot({\mathbf x}-{\mathbf x}')}.
 \end{equation}
By applying the fermionic anti-periodic boundary condition $G_\nu(t_E,t_E') = -G_\nu(t_E+\beta_R,t_E')$, the thermal propagator $G_\nu(x_E,x_E')$ can be constructed in a manner analogous to the scalar case:
 \begin{equation}
 \begin{split}
       G_{\nu}&=\sum_n (-1)^nG_E(t_E-t_E'+\beta_Rn)\\
       &=\int^\infty_0\mathrm{d}\omega\int\frac{\mathrm{d}^2{\mathbf k}}{(2\pi)^2}\frac{\sinh(\beta_R \omega/2-\omega|t_E-t_E'|)}{\cosh(\beta_R\omega/2)}\frac{\sinh\pi(\omega+i \gamma^{\hat{0}} \gamma^{\hat{3}}/2)}{\pi^2}K_{i\omega- \gamma^{\hat{0}} \gamma^{\hat{3}}/2}(m_\perp\xi)K_{i\omega- \gamma^{\hat{0}} \gamma^{\hat{3}}/2}(m_\perp\xi')e^{i{\mathbf k}\cdot({\mathbf x}-{\mathbf x}')}.
       \end{split}
 \end{equation}

The diagonal part of the propagator becomes:
\begin{equation}\label{Eq.FGv1}
    G_{\nu}(x_E,x_E)=\sum_{s=\pm}\int^\infty_0\mathrm{d}\omega\int\frac{\mathrm{d}^2{\mathbf k}}{(2\pi)^2}\frac{\sinh({\beta_R} \omega/2)}{\cosh({\beta_R}\omega/2)}\frac{i s \cosh\pi \omega}{\pi^2}K^2_{i\omega- s/2}(m_\perp\xi).
\end{equation}
As before, we consider the chiral limit and focus on the phase boundary where $m=0$,
\begin{equation}\label{eq:ftrg}
\begin{aligned}
    G_{\nu}(x_E,x_E;m=0) &= \int_0^\infty \mathrm{d}\omega \frac{\omega}{\pi^2 \xi^2} \tanh({\beta_R} \omega/2) \\
    &= \int_0^\L \mathrm{d}\o\frac{\o}{\pi^2 a^2\xi^2}+\int_0^\infty \ud\o \frac{\o}{\pi^2 a^2 \xi^2} \frac{2}{e^{\beta \o}+1}\\
    &=\frac{1}{6a^2 \xi^2}\left(\frac{3\Lambda^2}{ \pi^2}-T^2\right).
\end{aligned}
\end{equation}
In the second line, we have performed a variable substitution $\o \to \o/a$ and introduced an acceleration-independent cutoff $\Lambda$ to regularize the divergent term (the first term), which corresponds to the Rindler vacuum contribution, $G_{\nu=0}(x_E,x_E;m=0)$.
Substituting this result into the gap equation Eq.~(\ref{eq:gap simple}) and using the properties that the trace of odd-number gamma matrices vanishes, we have
\begin{equation}
    \frac{1}{G_\pi}=\frac{1}{6a^2 \xi^2}\left(\frac{3\Lambda^2}{ \pi^2}-T^2\right)
\end{equation}
Therefore, we get the critical temperature:
\begin{equation}\label{eq:Tcferimion}
    T_c(\x) = a\xi\sqrt{\frac{3\Lambda^2}{\pi^2}-\frac{6}{G_\pi}}.
\end{equation}
This expression agrees with the result in Ref.\cite{Ohsaku:2004rv}.
Furthermore, by substituting the accelerated observer’s worldline condition $\xi=1/a$, one finds that the critical temperature is independent of the acceleration $a$, $T_c(\x=1/a)=\sqrt{{3\Lambda^2}/{\pi^2}-{6}/{G_\pi}}\equiv T_{c0}$. As discussed in Section~\ref{c2}, renormalization schemes are not unique. In Sec.~\ref{ap0}, we present a supplemental alternative scheme that yields results different from those in this subsection and in Sec.~\ref{secnlsm}.

\begin{figure}[tp]
\begin{minipage}[t]{0.45\linewidth}
\includegraphics[width=1\columnwidth]{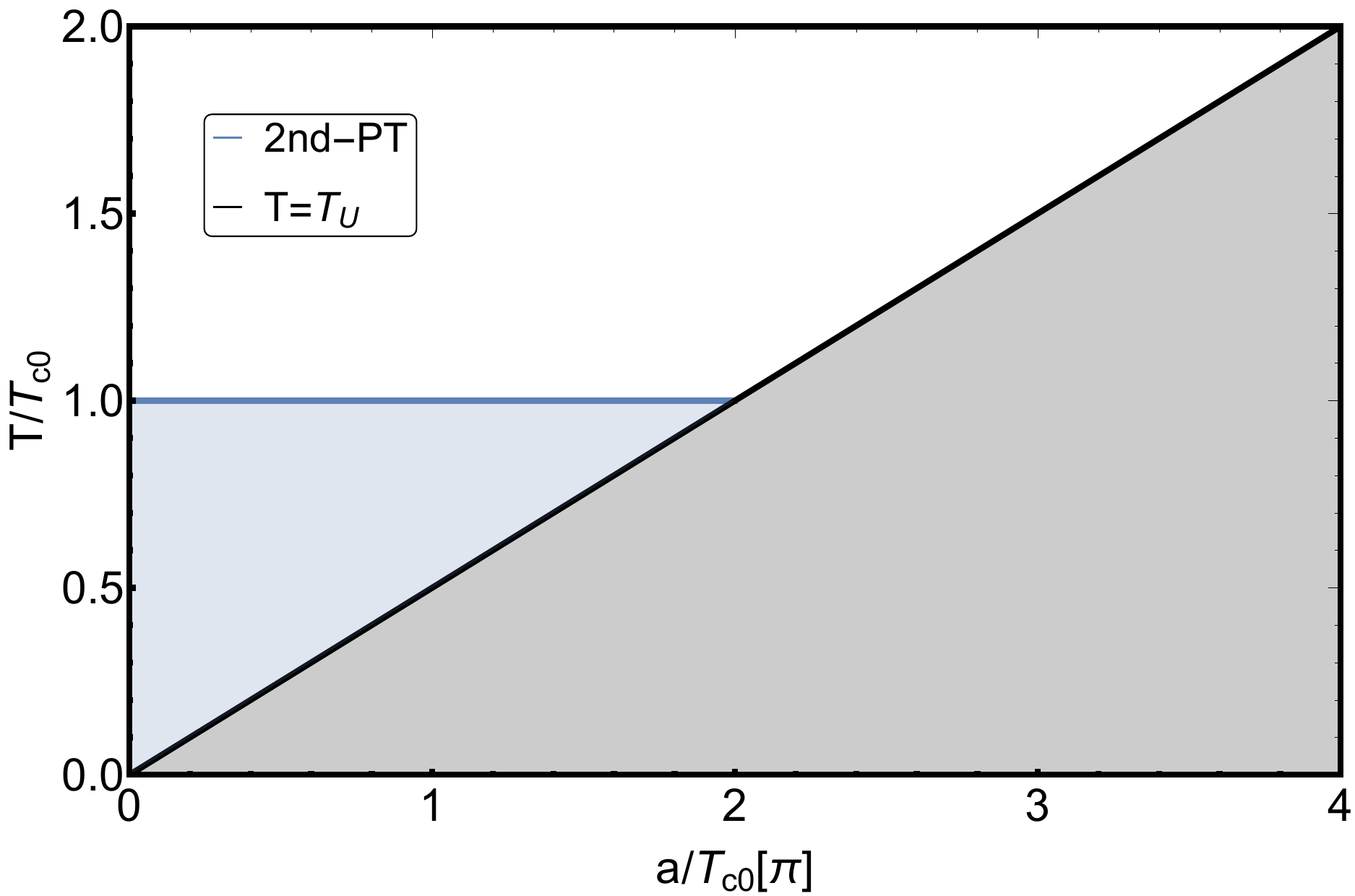}
    \caption{The $T-a$ phase diagram based on Eq.~(\ref{eq:Tcboson})(for bosons) and Eq.~(\ref{eq:Tcferimion})(for fermions). $T_{c0}$ denotes the critical the critical temperature in $a=0$. Blue region denotes the chiral symmetry breaking phase, while the white region denotes the chiral symmetry restored phase. Grey region denotes area of $T<T_U$ which is not explored in this study.}
    \label{fig:T-aphase-0}
\end{minipage}%
\hfill
\end{figure}

Figure~\ref{fig:T-aphase-0} shows the phase diagram calculated from Eqs.~\eqref{eq:Tcboson} and \eqref{eq:Tcferimion}.
The blue region corresponds to the phase with spontaneously broken chiral symmetry, the white region indicates restored symmetry, and the grey region marks the unphysical domain where $T<T_U$.
Although it may be intuitive to expect acceleration-induced thermal effects to lower the critical temperature, our results reveal that the critical temperature is, in fact, independent of acceleration, as along as $T_{c0}>T_U$. 
As shown in Eqs.~\eqref{eq:Tcboson} and \eqref{eq:Tcferimion}, the phase transition is governed solely by the temperature perceived by an accelerated observer with respect to the Rindler vacuum. Acceleration manifests its influence on phase transitions only indirectly via the temperature dependence. For instance, if one assumes the accelerated observer's temperature to be fixed to the Unruh relation $T=T_U=a/2\pi$, as has been considered in Refs.~\cite{Ohsaku:2004rv,Ebert:2006bh,Castorina:2012yg,Casado-Turrion:2019gbg,Basu:2023bcu}, one can determine a critical acceleration $a_c=2\pi T_c$, above which($a>a_c$) chiral symmetry is restored.


\subsection{Another renormalization scheme}\label{ap0}
In Secs.~\ref{secnlsm} and~\ref{subsc.njl}, we have adopted the renormalization scheme corresponding to the choice of the Rindler vacuum. However, as discussed in Sec.\ref{c2}, in the presence of acceleration the vacuum state is not unique. In this subsection, we present the explicit formulation of an alternative renormalization scheme in accordance to the choice of the Minkowski vacuum [see Eq.~(\ref{eq:re2})].

We consider the bosonic case first. In Sec.~\ref{secnlsm}, we regularize $G_\n(x_E, x_E;m=0)$ by subtracting its Rindler vacuum counterpart $G_{\n=0}(x_E,x_E;m=0)=\lan 0_R|\f^2(x_E;m=0)|0_R\ran$ with $|0_R\ran$ the Rindler vacuum state. In this subsection, we renormalize $G_\n(x_E, x_E;m=0)$ by subtracting its Minkowski vacuum counterpart $G_{\n=1}(x_E,x_E;m=0)=\lan 0_M|\f^2(x_E;m=0)|0_M\ran$ with $|0_M\ran$ the Minkowski vacuum state. The result is 
\begin{equation}
	G^{ren}_\n(x_E, x_E;m=0)=G_\nu(x_E,x_E;m=0)-G_{\nu=1}(x_E,x_E;m=0)=\frac{\nu^2-1}{\xi^2}\frac{1}{48\pi^2}.
\end{equation}
Under this renormalization scheme, Eq.~(\ref{eq:sclar sigma}) is modified into
\begin{equation}
	\begin{aligned}
		\sigma^2
		&=f^2_\pi-\frac{1}{(a\xi)^2} \frac{N}{12}\left(T^2-\frac{a^2}{(2\pi)^2}\right).
	\end{aligned}
\end{equation}
Substituting the accelerating observer's world line $\xi=1/a$, it shows that the acceleration would enhance the chiral symmetry breaking~\cite{Chernodub:2025ovo}. Accordingly, the critical temperature is
\begin{equation}\label{eq.tce}
	\begin{aligned}
		T_c(\x=1/a)&=\sqrt{\frac{12 f^2_\pi}{N}+(\frac{a}{2\pi})^2} \\
		&=\sqrt{T_{c0}^2+(\frac{a}{2\pi})^2}.
	\end{aligned}
\end{equation}
This result is consistent with Ref.~\cite{Chernodub:2025ovo}. 

\begin{figure}[tp]
	\begin{minipage}[t]{0.45\linewidth}
		\includegraphics[width=1\columnwidth]{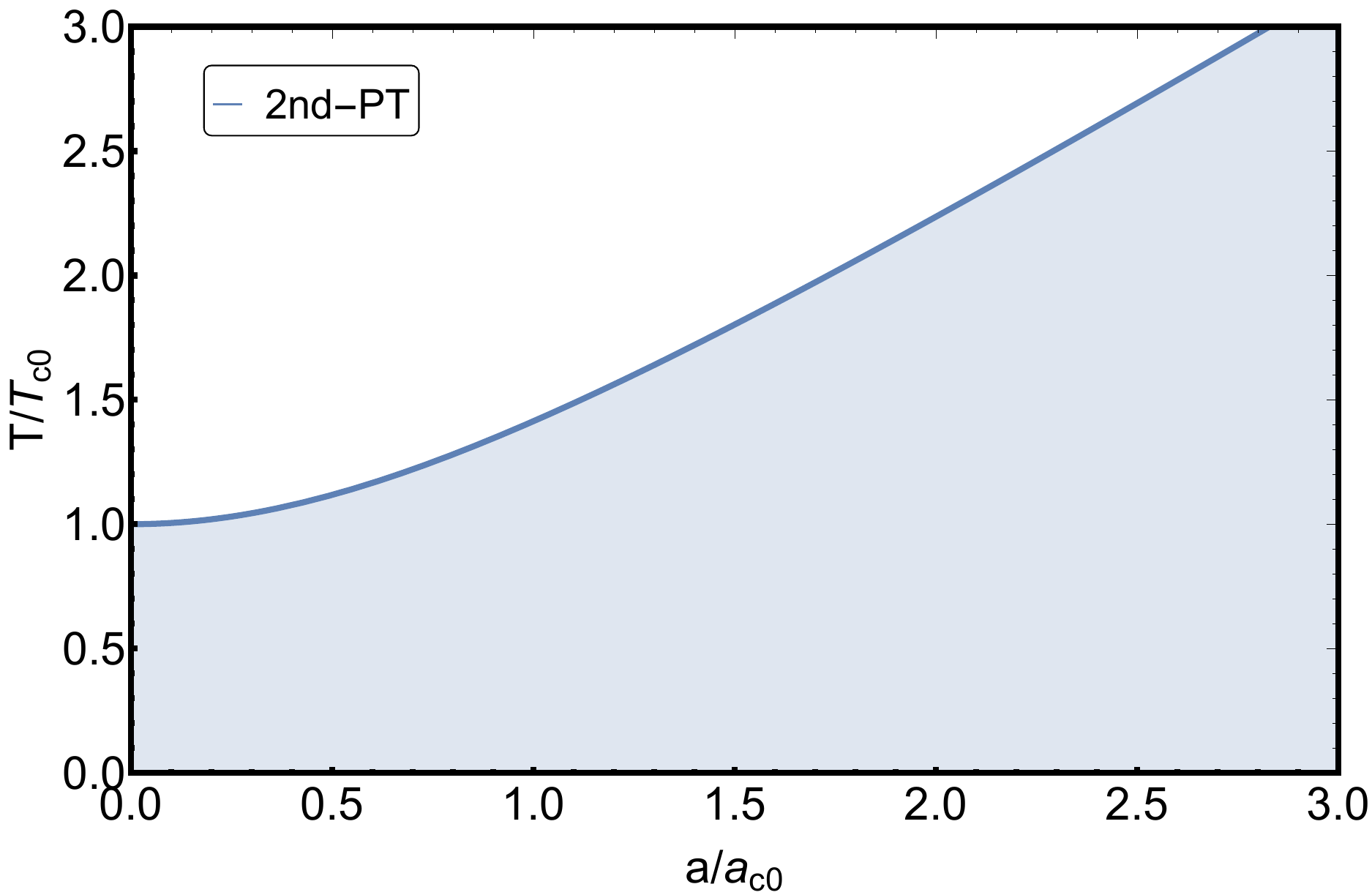}
		\caption{The $T-a$ phase diagram calculated from Eq.~(\ref{eq.tce}) and \eq{anotherfermion}.}
		\label{fig:Tc-a}
	\end{minipage}%
	\hfill
\end{figure}

Similarly, we can also apply the same renormalization scheme to the fermionic case. It is easily to verified that, the Eq.~(\ref{eq:ftrg}) is modified to
\begin{equation}
	G^{ren}_{\nu}(x_E,x_E;m=0)=\frac{1}{6a^2 \xi^2}\left(\frac{3\Lambda^2}{ \pi^2}-T^2+\frac{a^2}{(2\pi)^2}\right).
\end{equation}
And the critical temperature for the fermion case is
\begin{equation}\label{anotherfermion}
	\begin{aligned}
		T_c(\x=1/a)&=\sqrt{\frac{3\Lambda^2}{\pi^2}-\frac{6}{G}+\frac{a^2}{(2\pi)^2}} \\
		&=\sqrt{T_{c0}^2+(\frac{a}{2\pi})^2},
	\end{aligned}
\end{equation}
which coincide with \eq{eq.tce}. The corresponding phase diagram calculated from \eq{eq.tce} and \eq{anotherfermion} is shown in Fig.~\ref{fig:Tc-a}. 

\section{NJL model under acceleration and rotation} \label{c5}
In this section, we investigate NJL model under both acceleration and rotation. For simplicity, we assume that the direction of acceleration is aligned with the rotation axis. 
The spacetime metric describing such an accelerating and rotating frame is given by:
\begin{equation}
\begin{aligned}
        g_{00} &= \xi^2 - (\boldsymbol{\bar{\Omega}} \times \boldsymbol{x})^2, \\
        g_{0i} &= g_{i0} = -(\boldsymbol{\bar{\Omega}} \times \boldsymbol{x})^i, \\
        g_{ij} &= -\delta_{ij},
    \end{aligned}
\end{equation}
where $i,j=1,2,3$, with dimensionless angular velocity $\boldsymbol{\bar{\Omega}} = (0,0,\bar{\Omega})$ aligned with the acceleration axis and $\boldsymbol{x}=(x_1,x_2,\xi)$. The dimensionless angular velocity obeys $\bar{\Omega}={\O}/a$, where $\O$ is the usual energy-dimension angular velocity.
The vierbein $e^\mu_{\hat{m}}$ for acceleration and rotation are
\begin{equation}
     e^0_{\hat{0}}= \frac{1}{\xi}\,,\,e^i_{\hat{0}}=-\frac{(\boldsymbol{\bar{\Omega}} \times \boldsymbol{x})^i}{\xi} ,e^0_{\hat{j}}=0, e^i_{\hat{j}}=\delta^i_j.
\end{equation}
Accordingly, the spacetime-dependent gamma matrices are
\begin{equation}
\begin{gathered}
\gamma^0(x)=\frac{1}{\xi} {\gamma^{\hat{0}}},\quad
\gamma^i(x)=\frac{(\boldsymbol{\bar{\Omega}} \times \boldsymbol{x})^i}{\xi} {\gamma^{\hat{0}}}+\gamma^{\hat{i}},
\end{gathered}
\end{equation}
and the only nonzero spin connection is
\begin{equation}
\Gamma_0=-\frac{i}{2} \boldsymbol{\bar{\Omega}} \cdot \boldsymbol{\sigma}+\frac{1}{2} \gamma^{\hat{0}} \gamma^{\hat{3}},
\end{equation}
where $\boldsymbol{\sigma}$ denotes $4\times 4$ Pauli matrices. Under this setup, the NJL model action in the rotating and accelerating frame becomes
\begin{equation}
	\label{actioninar}
S=\int \ud^4 x\left\{\bar{\psi}\left[{\gamma^{\hat{0}}}\left(i \partial_0+\bar{\Omega}\left(\frac{\sigma_3}{2}+\hat{L}_z\right)+\frac{i}{2}  \gamma^{\hat{0}} \gamma^{\hat{3}}\right)+\xi i \gamma^{\hat{\imath}} \partial_i-m_0 \xi\right] \psi+\frac{G_\pi}{2} \xi\left[(\bar{\psi} \psi)^2+\left(\bar{\psi} i \gamma_5 \psi\right)^2\right]\right\},
\end{equation}
with $\hat{L_z}=x_1(-i \partial_2)-x_2(-i \partial_1)$ as the orbital angular momentum operator. The Dirac equation consequently takes the form:
\begin{equation}\label{eq:dirac}
{\left[\frac{\gamma^{\hat{0}}}{\xi}\left(i \partial_0+\bar{\Omega}\left(\frac{\sigma_3}{2}+\hat{L}_z\right)+\frac{i}{2}  \gamma^{\hat{0}} \gamma^{\hat{3}}\right)+i \gamma^{\hat{\imath}} \partial_i-m\right] \psi=0}.
\end{equation}
Following the same procedure in the pure-acceleration case, we can derive the squared Dirac operator $\hat{D}^2$:
\begin{equation}\label{eq:A2accrot}
\hat{D}^2=\frac{1}{\xi^2}\left\{\left[i \partial_0+\left(\frac{\sigma_3}{2}+\hat{L}_z\right) \bar{\Omega}\right]^2-\frac{1}{4} \right\}+\partial_3^2+\frac{1}{\xi} \partial_3+{\gamma}^{\hat{0}} {\gamma}^{\hat{3}} \frac{1}{\xi^2} i\left[i \partial_0+\left(\frac{\sigma_3}{2}+\hat{L}_z\right)\bar{\Omega}\right]+\partial_1^2+\partial_2^2.
\end{equation}
By analogy with the projection operator used in the purely accelerating case, we now also introduce the rotational projection operator $S^\pm=(1\pm i\gamma^{\hat{1}} \gamma^{\hat{2}} )/2$ to account for the spin structure induced by rotation. Therefore, the full propagator can be decomposed as:
\begin{equation}
\begin{gathered}
        G=\sum_{s_1,s_2=\pm 1} \frac{1}{2} (1+s_1 \gamma^{\hat{0}} \gamma^{\hat{3}} ) \frac{1}{2} (1+s_2 i \gamma^{\hat{1}} \gamma^{\hat{2}} )  \mathcal{G}^{s_1,s_2},
\end{gathered}
\end{equation}
where
\begin{equation}
\begin{gathered}
        \begin{aligned}
                    \mathcal{G}^{s_1,s_2}=\frac{i}{\b} &\sum_{l,k,n} \int_0^\infty\ud\omega  \frac{1}{\left( {\o_n-i\bar{\O}j +\frac{s_1}{2}} \right)^{2} - \left( {i\omega + \frac{s_1}{2}} \right)^{2}} \frac{(2is_1 \omega +1) \cosh (\pi \omega )}{\pi^2}K_{{i\omega} + \frac{s_1}{2}}\left( {m_\perp \xi_{1}} \right)K_{{i\omega} + \frac{s_1}{2}}\left( {m_\perp \xi_{2}} \right) \\
                    &\times \frac{1}{2\pi } \frac{1}{N_{l,k}}  e^{i(j-s_2/2) (\theta_1-\theta_2)} J_{j-s_2/2}(p_{l,k} r_1) J_{j-s_2/2}(p_{l,k} r_2),
        \end{aligned}
\end{gathered}
\end{equation}
with cylindrical coordinates $r=\sqrt{x^2+y^2}$, transverse mass $m_\perp=\sqrt{m^2+p_{l,k}^2}$, angular quantum number $j=l+1/2$ ($l$ integer), $J_\n$ the Bessel function of the first kind, and $p_{l,k}$ is the $k$-th root of $J_l(p R)$ with $R$ the transverse spatial boundary of the system. We adopt the same boundary condition as Ref.~\cite{Ebihara:2016fwa}, namely the vanishing normal flux at the spatial boundary $r=R$ with causality constraint $\O R<1$. This choice determines the normalization factors:
\begin{equation}
N_{l,k}=\begin{cases}
\frac{J^2_{l+1}(p_{l,k}R)R^2}{2} & l\geq0 \\
\frac{J^2_{l}(p_{l,k}R)R^2}{2} & l<0
\end{cases}.
\end{equation}

Hereafter, we focus on the case $T=T_U$. Repeating the computational procedure analogous to the acceleration case and taking the chiral limit, we derive the only non-trivial gap equation :
\begin{equation}\label{eq:gapm=0}
\begin{aligned}
  \frac{1}{G_\pi}=&\sum_{l,k,s_1} \int_0^\infty \ud\omega \, \frac{1}{2\pi} \frac{1}{N_{l,k}} \, \frac{-is_1}{2a} \frac{\cosh(\frac{\pi \omega}{a})}{\pi^2} \left\{\tanh \left(\frac{\omega-{\Omega} j}{2 T}\right)+\tanh \left(\frac{\omega+{\Omega} j}{2 T}\right)\right\} \\
  & \times K^2_{i\frac{\o}{a} +s_1 \frac{1}{2}}\left( {m_\perp \xi} \right) \left[ J^2_{l}(p_{l,k} r)+J^2_{l+1}(p_{l,k} r) \right],
\end{aligned}
\end{equation}
where we have adopted variable substitution $\o \to \o/a$. 

Equation~(\ref{eq:gapm=0}) determines the constituent quark mass (i.e., the chiral condensate) as a function of either the angular velocity $\Omega$ or the acceleration $a$.
The numerical results for $m(a)$ and $m(\Omega)$ are shown in Fig.~\ref{fig:n=1,m0=0,m-a} and Fig.~\ref{fig:n=1,m0=0,m-w}, respectively, using a cutoff $\Lambda = 1$ GeV, $G_\pi = 22/\Lambda^2$, and setting $r = 0$, $\xi=1/a$. We approximate the discrete momentum sum by a continuous integral, which is a valid simplification at the center point where r=0. These results indicate that both acceleration and rotation enhance chiral symmetry restoration. This behavior can be attributed to the fact that acceleration induces an effective temperature, while rotation generates an effective chemical potential, both of which tend to suppress the chiral condensate.
Figure~\ref{fig:n=1,m0=0,m-a} shows that chiral symmetry is restored at $a \approx 1.05$ GeV in the absence of rotation, slightly below the analytical critical value $a_c = 1.11$ GeV, likely due to numerical accuracy.  Figure~\ref{fig:n=1,m0=0,m-w} indicate a first order phase transition along with increasing $\O$: When $\O$ increase, the gap equation exist multiple solutions which lead to the constituent quark mass vanish discontinuously.

\begin{figure}[htbp]
\begin{minipage}[t]{0.45\linewidth}
\includegraphics[width=1\columnwidth]{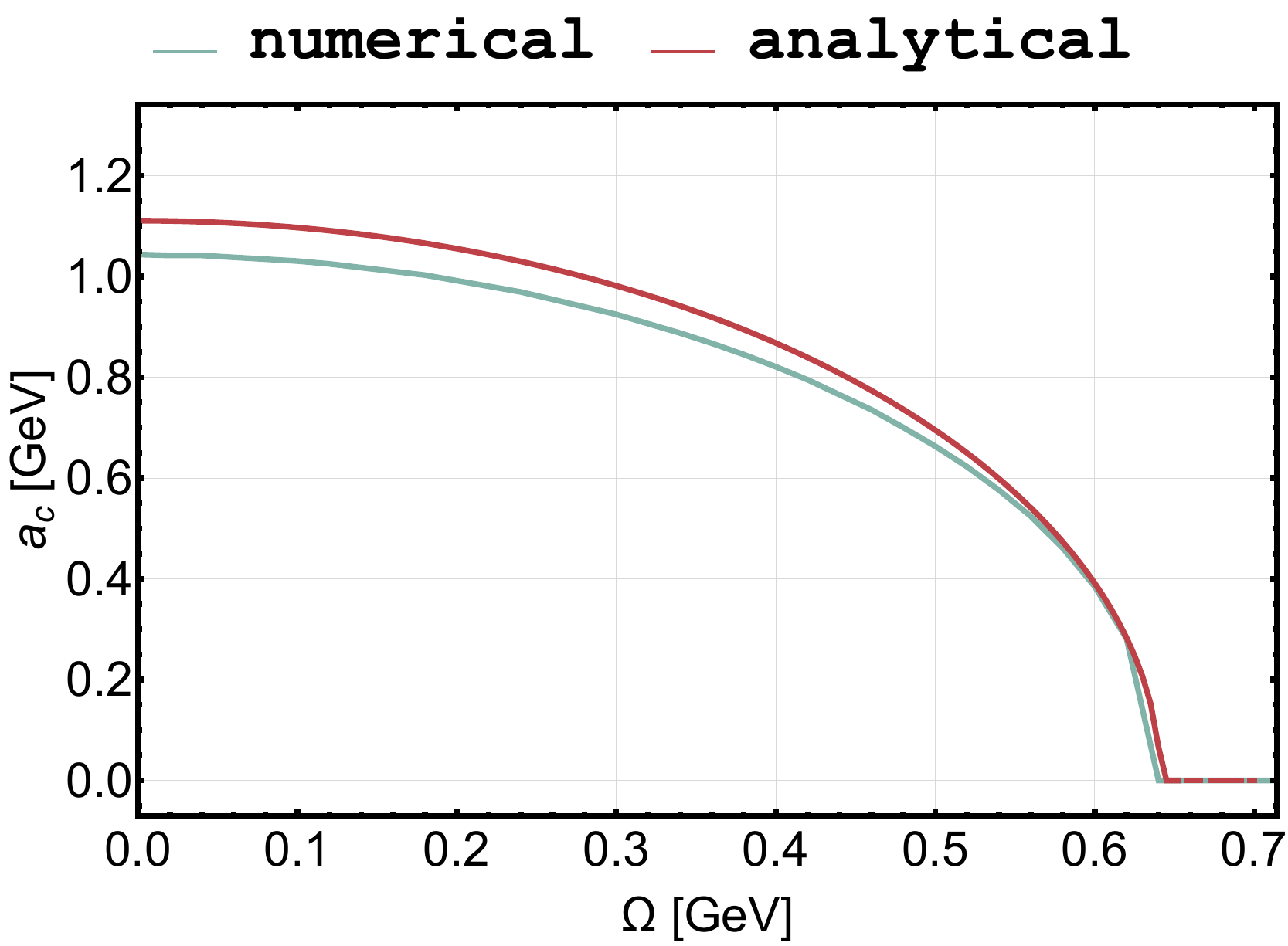}
    \caption{Critical acceleration $a_c$ as a function of $\Omega$ at $r=0$ and $T=T_U$ calculated from numerical solving Eq.~(\ref{eq:gapm=0}) and analytical equation Eq.~(\ref{eq:analyticalgap}). The results of both are consistent with each other.}
    \label{fig:ac-w-1}
\end{minipage}%
\hfill
\begin{minipage}[t]{0.45\linewidth}
\includegraphics[width=1\columnwidth]{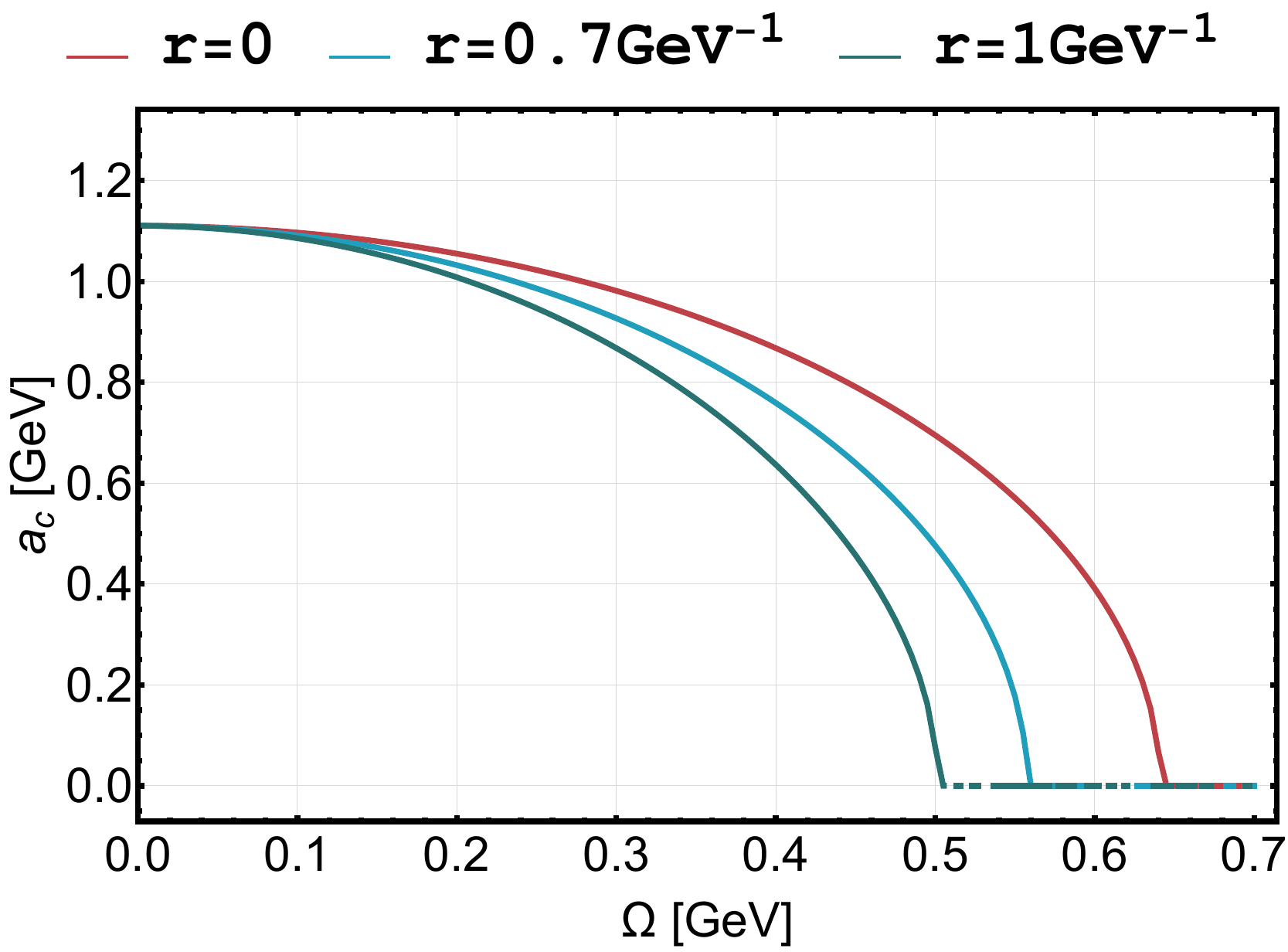}
    \caption{Critical acceleration $a_c$ for $T=T_U$ as a function of $\Omega$ with different radius $r$, calculated from Eq.~(\ref{eq:analyticalgap}). Rotation suppresses the critical acceleration. The rotation effect becomes more significant with a larger radius.}
    \label{fig:ac-w-2}
\end{minipage}
\end{figure}

 The gap equation can be decomposed into a rotation-independent part $F_0$ and a rotation-dependent part $F_\Omega$:
\begin{equation}
F_0+ F_\Omega = \frac{1}{G_\pi},
\end{equation}
where
\begin{equation}\label{eq:gap with a and omega}
\begin{aligned}
F_0&=\sum_l\int_0^\infty \ud\o \int \frac{\ud^2 p_t}{(2\pi)^2}  \frac{2\cosh(\frac{\pi \omega}{a} )}{a\pi^2}
       \im\left[K^2_{{i\frac{\o}{a}} + \frac{1}{2}}\left( {m_\perp \xi} \right) \right] \left [ J^2_{l}(p_t r)+J^2_{l+1}(p_t r) \right],\\
F_\Omega&=-\sum_l\int_0^\infty \ud\o \int_0^\infty \frac{p_t \ud p_t}{2\pi} \frac{2 \cosh(\pi \o /a)}{a \pi ^2}\left[ \frac{1}{e^{\beta(\o+\O j)}+1}+\frac{1}{e^{\beta(\o-\O j)}+1} \right]\im\left[K^2_{\frac{i\omega}{a} + \frac{1}{2}}\left( {m_\perp \xi} \right) \right] \left [ J^2_{l}(p_t r)+J^2_{l+1}(p_t r) \right].
\end{aligned}
\end{equation}
In Eq.~(\ref{eq:gap with a and omega}), the momentum summation for $k$ has been approximated by the integration of $p_t$ valid for $r \ll R$.
By setting $m=0$, $T=T_U$, and $\xi=1/a$, the critical acceleration $a_c$ can be evaluated analytically as a function of $\Omega$. We find:
\begin{equation}
F_0=\frac{\Lambda^2}{2\pi^2},
\end{equation}
as derived previously. And the rotation-dependent term becomes
\begin{equation}
F_\Omega=-\frac{a^2 \left(r^4 \Omega^4 +1\right)-r^2 \Omega^4 +3\Omega^2}{24\pi^2 \left(r^2 \Omega^2 -1\right)^2},
\end{equation}
with detailed derivation provided in Appendix.~\ref{ap2}. We thus obtain a simplified form of the gap equation:
\begin{equation}\label{eq:analyticalgap}
G_\pi\left(\frac{\Lambda^2}{2\pi^2}-\frac{a^2 \left(r^4 \Omega^4 +1\right)-r^2 \Omega^4 +3\Omega^2}{24\pi^2 \left(r^2 \Omega^2 -1\right)^2}\right)=1.
\end{equation}
The critical acceleration $a_c$ as a function of $\Omega$ for $r=0$ is shown in Fig.~\ref{fig:ac-w-1}, obtained by solving Eq.~(\ref{eq:analyticalgap}) analytically and Eq.~(\ref{eq:gap with a and omega}) numerically. The agreement confirms the validity of our analytical result. The monotonic decrease of $a_c$ with increasing $\Omega$ indicates enhanced symmetry restoration by rotation, consistent with results from effective models for purely rotating case~\cite{Chen:2015hfc,Jiang:2016wvv,Chen:2021aiq}. Figure~\ref{fig:ac-w-2} demonstrates the radius-dependent behavior of $a_c(\Omega)$. Larger radii enhance rotation effect: $a_c$ approaches zero at smaller $\Omega$ values compared to central regions, showing enhanced rotational effects away from the system's center.
\begin{figure}[tp]
\begin{minipage}[t]{0.45\linewidth}
\includegraphics[width=1\columnwidth]{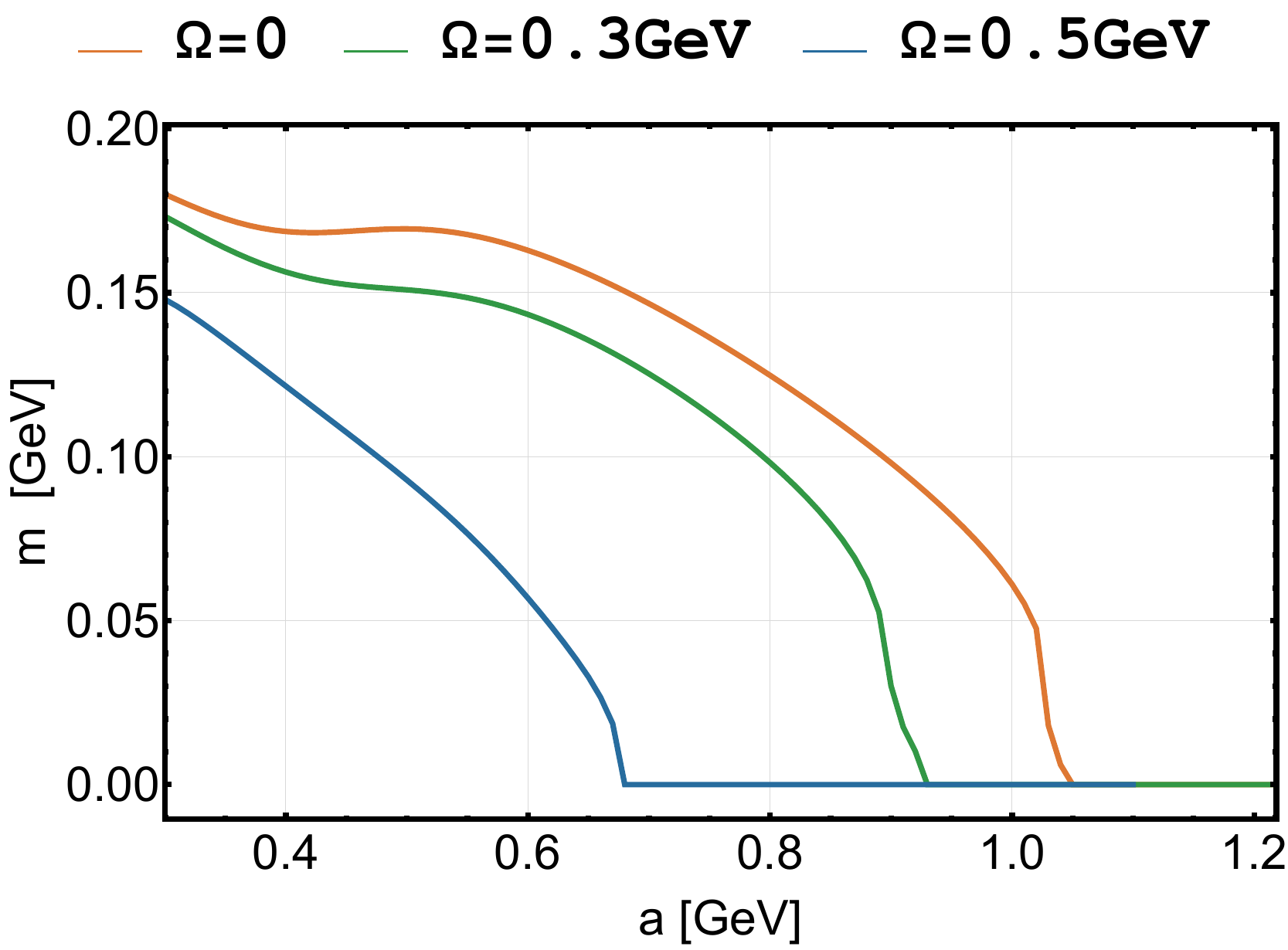}
    \caption{The constituent quark mass as a function of acceleration $a$ with $T=T_U$ at different $\Omega$, calculated from Eq.~(\ref{eq:gapm=0}). }
    \label{fig:n=1,m0=0,m-a}
\end{minipage}%
\hfill
\begin{minipage}[t]{0.45\linewidth}
\includegraphics[width=1\columnwidth]{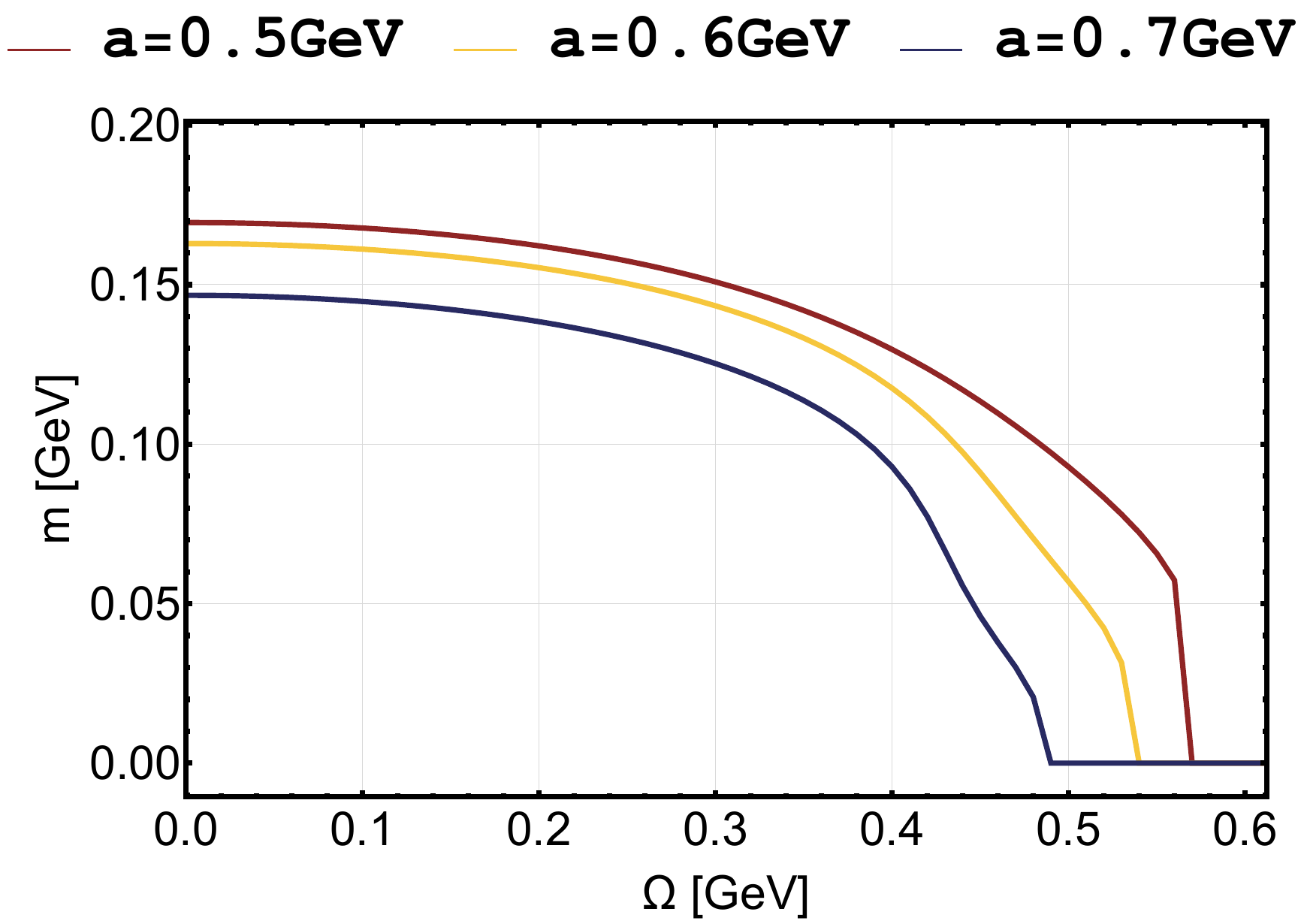}
    \caption{The constituent quark mass as a function of $\Omega$ for different acceleration $a$ with $T=T_U$, calculated from Eq.~(\ref{eq:gapm=0}). }
    \label{fig:n=1,m0=0,m-w}
\end{minipage}
\end{figure}

\section{summary}\label{c6}
In this study, we investigate chiral symmetry breaking behavior for accelerating and rotating systems using effective models in non-inertial frames. We solve the Klein-Gordon and Dirac equations in an accelerating frame and construct the phase diagram in the $T$–$a$ plane.
Our analysis confirms that the choice of renormalization scheme can affect the physical implications of acceleration on chiral symmetry breaking. Using the renormalization scheme in Eq.~(\ref{eq:re1}), we find that phase transitions depend solely on the temperature $T$ as measured by the observer as long as $T_U<T$.
The role of acceleration is encoded through its relation to $T$. In particular, when the Unruh relation $T = a/2\pi$ is imposed, acceleration induces chiral symmetry restoration via effective thermalization. On the other hand, using the renormalization scheme as given in Eq.~(\ref{eq:re2}), similar to Ref.~\cite{Chernodub:2025ovo}, an acceleration induced enhancement of chiral symmetry breaking is obtained.

We further extend our analysis to frames with both rotation and acceleration. Two restoration mechanisms are identified: one from the thermal effect induced by acceleration (at fixed temperature $T=T_U$), and the other from the rotation-induced modification of the ``chemical potential", consistent with previous findings (e.g., Ref.~\cite{Chen:2021aiq} for a review). We further define the critical acceleration $a_c$ and derive its dependence on the angular velocity $\Omega$. Analytical solutions to the gap equation reveal that increasing $\Omega$ reduces $a_c$, and this suppression becomes more pronounced at larger radial distances.

Several open issues remain unresolved. While our calculations clearly demonstrate that different renormalization schemes—whether defined with respect to the Minkowski vacuum or the Rindler vacuum—can lead to dramatically different physical conclusions, the physical interpretation of these results and their interrelationships remain obscure. This ambiguity underscores the need for further investigation. 
In this work, the rotation and acceleration are aligned in the same direction to simplify the discussion. For non-aligned case, the calculation would be much more complicated. For the case of free fermions, the work of Ref~\cite{Palermo:2021hlf} has indicates that the pressure depends on a term proportional to $(a\cdot \O)^2$. It can therefore be expected that the effects of rotation and acceleration are influenced by their relative orientation.  However, the scenario in which they possess perpendicular components is considerably more complex and falls outside the scope of the present study.
We leave the resolution of these fundamental issues, along with the exploration of deconfinement transitions under acceleration, to future study.

\section*{Acknowledgment}
We thank Kenji Fukushima and Ji-Chong Yang for helpful discussions. This work is supported by the Natural Science Foundation of Shanghai (Grant No. 23JC1400200), the National Natural Science Foundation of China (Grant No. 12225502 and No. 12147101), and the National Key Research and Development Program of China (Grant No. 2022YFA1604900).

\appendix
\section{Accelerating frame and Rindler coordinates}\label{apprind}
We give a brief discussion about the accelerating frame and Rindler coordinates in this appendix. More discussions can be found in, e.g., Ref.~\cite{Birrell:1982ix}. Readers who are familiar with Rindler coordinates can omit this appendix. 

Consider a particle moving in the Minkowski spacetime $\cal M$ with coordinates $x_M^\m=(t_M,\bx_M)$~\footnote{In this appendix, we use a subscript $M$ to distinguish Minkowski coordinates.}. Its four-velocity is,
\begin{eqnarray}
	\label{eq:veloinM}
	u_{M}^\m=\frac{d x^\m_M}{d\t}=\g_M(1,\bv_M),
\end{eqnarray}
with $\t$ the proper time, $\g_M=dt_M/d\t$ the Lorentz factor, and $\bv_M=d\bx_M/dt_M$ the three-velocity of the particle. Note that $u^2_M=1$. The four-acceleration of the particle is 
\begin{eqnarray}
	\label{eq:accinM}
	\a_{M}^\m=\frac{d u^\m_M}{d\t}=\g^2_M(\g_M^2\bv_M\cdot\bm a_M,\bm a_M+\g_M^2\bv_M\cdot\bm a_M\bv_M),
\end{eqnarray}
where $\bm a_M=d^2\bx_M/dt_M^2$ is the three-acceleration of the particle. Note that $u_M\cdot\a_M=0$ so that $\a_M^\m$ is space-like, $\a_M^2=-a^2<0$. Suppose the particle is moving along positive $z_M$ direction. We can thus focus only on the two dimensional spacetime $(t_M, z_M)$. It is easy to show that $\g_M^3 a^z_M =a$ and $\a^z_M = \g_M a$ with $a>0$. Therefore,
\begin{eqnarray}
	\label{eq:accinMccdd}
	a=\frac{d u^z_M}{dt_M}.
\end{eqnarray}
Because $\a^2_M$ is a Lorentz scalar (so does $a$), we can go into the instantaneous inertial frame comoving with the particle at each instant where $a$ is just the acceleration measured in that reference frame. For this reason, $a$ is called the proper acceleration of the particle. 

Let us find out the trajectory in Minkowski coordinates of a particle under constant proper acceleration $a>0$. From \eq{eq:accinMccdd}, we find that $\g_M v^z_m=a t_M$ (supposing that $v_M^z=0$ at $t_M=0$) and thus
\begin{eqnarray}
	\label{eq:trajectory}
	v^z_M(t_M)=\frac{at_M}{\sqrt{1+(at_M)^2}},
\end{eqnarray}\
and $\g_M(t_M)=\sqrt{1+(at_M)^2}$. Integrating $v_M^z(t_M)$ over $t_M$, we obtain 
\begin{eqnarray}
	\label{eq:tdddrajectory}
	z_M(t_M)=\frac{1}{a}\sqrt{1+(at_M)^2}+z_M(0)-\frac{1}{a}.
\end{eqnarray}
The particle is initially at $z_M(0)$ and then accelerated towards the positive $z_M$ direction (such that $z_M(t_M)\geq z_M(0)$). The trajectory is half of a hyperbola (another half of the hyperbola corresponds to a particle under the same proper acceleration $a$ but moving in $-z_M$ direction),
\begin{eqnarray}
	\label{eq:tdddrajectorydd}
	t_M^2-\lb z_M-z_M(0)+\frac{1}{a}\rb^2=-\frac{1}{a^2},\quad z_M>z_M(0).
\end{eqnarray}

Now suppose that the particle is an observer, so that its reference frame is the accelerating frame associated with the particle. Its path is the above hyperbola in the Minkowski coordinates. Evidently, it is more convenient to make a coordinate transformation from $(t_M, z_M)$ to $(\t,z)$ so that the observer is static in this coordinate system. Let
\begin{eqnarray}
	\label{eq:accoridad}
	t_M&=&\lb z+\frac{1}{a}\rb\sinh(a\t),\\
	z_M&=&\lb z+\frac{1}{a}\rb\cosh(a\t)+z_M(0)-\frac{1}{a}.
\end{eqnarray}
We have $t_M^2-(z_M-z_M(0)+1/a)^2=-(z+1/a)^2$. Thus, the observer at $z=0$ in the new coordinates $(\t, z)$ is accelerated by a proper acceleration $a$ and $\t$ is its proper time. An observer at an arbitrary $z$ is accelerated by a proper acceleration $a/(1+az)$. The line element $ds^2$ is then
\begin{eqnarray}
	\label{eq:lineaccoridad}
	ds^2=dt_M^2-dz_M^2=(1+az)^2 d\t^2-dz^2.
\end{eqnarray}
This coordinates is called Rindler coordinates. It is important to notice that $(\t, z)$ does not cover the whole Minkowski spacetime. In fact, for $z>-1/a$, $z_M-z_M(0)+1/a>|t_M|$ (called the right Rindler wedge ), while for $z<-1/a$, $z_M-z_M(0)+1/a<-|t_M|$ (called the left Rindler wedge). At $z=-1/a$, we have $z_M-z_M(0)+1/a=\pm t_M$ which are two light cones called Rinder horizons. Sometimes it is more convenient to use another coordinates $(t,\x)$ with $t=a\t, \x=z+1/a$, so that 
\begin{eqnarray}
	\label{eq:dlineaccoridad}
	ds^2=\x^2 d t^2-d\x^2.
\end{eqnarray}
The observer at $\x$ has constant proper acceleration $1/\xi$. This is the coordinates in \eq{eq.Rindler metric}.

\section{Chiral limit Gap equation under acceleration and rotation}\label{ap2}
In this appendix, we will provide a detailed derivation process for Eq.~(\ref{eq:analyticalgap}).
We start from Eq.~(\ref{eq:gap with a and omega}), where
\begin{equation}
\begin{aligned}
        F_0=\int_0^\infty \ud\o\int \frac{\ud^2 p_t}{(2\pi)^2}  \frac{1}{a} \frac{4\cosh(\pi \omega / a)}{\pi^2}\im\left[K^2_{\frac{i\omega}{a} + \frac{1}{2}}\left( {m_\perp \xi} \right) \right]=\frac{\Lambda^2}{2\pi^2},
\end{aligned}
\end{equation}
and
\begin{equation}
    F_\Omega=-\sum_l\int_0^\infty \ud\o \int_0^\infty \frac{p_t \ud p_t}{2\pi} \frac{2 \cosh(\pi \o /a)}{a \pi ^2}\left[ \frac{1}{e^{\beta(\o+\O j)}+1}+\frac{1}{e^{\beta(\o-\O j)}+1} \right]\im\left[K^2_{\frac{i\omega}{a} + \frac{1}{2}}\left( {m_\perp \xi} \right) \right] \left [ J^2_{l}(p_t r)+J^2_{l+1}(p_t r) \right] ,
\end{equation}
where we have used a variable substitution $\omega\to\omega/a$.
We handle rotational effects through analytic continuation from real to imaginary angular velocity ($\Omega = i\Omega_I$). This yields the $F_\Omega$ as:
\begin{equation}
\begin{aligned}
        F_\Omega= \int_0^\infty \ud\omega \int_0^\infty \frac{p_t \ud p_t}{2 \pi} \frac{2 \cosh(\pi \omega /a)}{a \pi ^2} \sum_{n=1}^\infty (-1)^n e^{-n \beta \omega} \im\left[K^2_{\frac{i\omega}{a} + \frac{1}{2}}\left( {m_\perp \xi} \right) \right] 4J_0(p_t r \sqrt{2-2\cos(n\beta \Omega_I)}) \cos\left(\frac{n \beta \Omega_I}{2}\right),
\end{aligned}
\end{equation}
where the Fermi-Dirac distribution $1/(\ue^{\beta (\omega \pm \Omega j)}+1)$ has been expanded via $-\sum_{n=1}^\infty (-1)^n \ue^{-n \beta (\omega \pm i\Omega_I j)}$. The angular momentum summation over $l$ is resolved through Bessel function identities\cite{Chen:2023cjt}:
\begin{equation}
    \sum_{n \in \mathbb{Z}} \mathrm{e}^{i n \theta} J_n^2(x)=J_0(x \sqrt{2-2 \cos \theta}).
\end{equation}
We can deal with the $K_{\frac{i\omega}{a} + \frac{1}{2}}\left( {m_\perp \xi} \right)$ by using\cite{candelas1976quantum}
\begin{equation}
    K_{i\nu}(\mu \xi)K_{i\nu}(\mu \xi')=\frac{1}{2} \int_{-\infty}^{\infty} \ud\lambda \ue^{i \nu \lambda} K_0(\mu \gamma_1),
\end{equation}
where $\g^2_1=\xi^2 + {\xi'}^2 +2\xi \xi' \cosh(\lambda)$. Thus
\begin{equation}
\begin{aligned}
        \im\left[K^2_{\frac{i\omega}{a} + \frac{1}{2}}\left( {m_\perp \xi} \right) \right]=-\frac{1}{4}i\left[ \int^\infty_{-\infty} \ud\l  \exp({i\frac{\o}{a} \l }) \left(2 \sinh(\frac{\l}{2})\right) K_0\left(m_\perp \xi \sqrt{2+2\cosh(\l)}\right)\right].
\end{aligned}
\end{equation}
To perform the $p_t$ integration, we expand $K_\nu$ by using the integral representation
\begin{equation}
    K_\nu(t)=\frac{1}{2} (\frac{1}{2}t)^\nu \int_0^\infty \exp \left(-z-\frac{t^2}{4z} \right) \frac{\ud z}{z^{\nu+1}}.
\end{equation}
Applying this formula and setting $\xi=1/a$, we have
\begin{equation}
\begin{aligned}
        &\int_0^\infty \frac{\ud z}{2z} e^{-\frac{1}{2} z} \int_0^\infty \frac{p_t \ud p_t}{2\pi} \exp\left[{-\frac{1+\cosh(\l)}{z} \frac{m^2+p_t^2}{a^2}}\right]J_0(p_t r \sqrt{2-2\cos(n\b \O_I)}) \\
        =& \int_0^\infty \frac{\ud z}{2z} e^{-\frac{1}{2} z}  \int_0^\infty\frac{p_t \ud p_t}{2\pi} \exp\left[-\frac{A}{z}(m^2+p_t^2)\right]J_0(p_tB)
        =\int_0^\infty \frac{\ud z}{2z} e^{-\frac{1}{2} z}\frac{z}{2\pi} exp\left[-\frac{zB^2}{4A}-\frac{Am^2}{z}\right]/(2A) \\
        =& \frac{1}{2\pi} \frac{m}{\sqrt{(2+2\cosh(\l))/a^2+r^2(2-2\cos(n\b\O_I))}}K_1(m\sqrt{(2+2\cosh(\l))/a^2+r^2(2-2\cos(n\b\O_I))}),
\end{aligned}
\end{equation}
where auxiliary parameters $A=\frac{1+\cosh(\lambda)}{a^2}$ and $B=r\sqrt{2-2\cos(n\beta \Omega_I)}$.

The integral over $\o$ can be evaluated using
\begin{equation}
    \int_0^{\infty} \ud\o \cosh(\frac{\pi\o}{a})\exp[-n\b \o+i\frac{\o}{a}\l]=\frac{i a}{2 (i a \beta  n+\lambda -i \pi )}+\frac{i a}{2 (i a \beta  n+\lambda +i \pi )}.
\end{equation}
Now $F_\Omega$ is simplified into
\begin{equation}\label{eq:ft1}
\begin{aligned}
        F_\Omega=&\frac{1}{\pi^3} \sum_{n=1} (-1)^n \int_{-\infty}^\infty \ud\lambda \left( \frac{1}{i a \beta  n+\lambda -i \pi }+\frac{1}{ i a \beta  n+\lambda +i \pi } \right)\frac{m}{\sqrt{(2+2\cosh(\l))/a^2+r^2(2-2\cos(n\b\O_I))}} \\
        & \times K_1(m\sqrt{(2+2\cosh(\l))/a^2+r^2(2-2\cos(n\b\O_I))}) \sinh(\frac{\l}{2}) \cos(\frac{n\b \O_I}{2}) \\
         =&\frac{1}{\pi^3} \sum_{n=1} (-1)^n \int_{C} \ud\lambda \left( \frac{1}{i a \beta  n+\lambda  } \right) i\cosh(\frac{\l}{2}) \frac{m}{\sqrt{(2-2\cosh(\l))/a^2+r^2(2-2\cos(n\b\O_I))}} \\
        & \times K_1(m\sqrt{(2-2\cosh(\l))/a^2+r^2(2-2\cos(n\b\O_I))})  \cos(\frac{n\b \O_I}{2}) ,
\end{aligned}
\end{equation}
where the contour $C$ is $-\infty-i\pi \to \infty -i\pi$  and $\infty+i\pi \to -\infty+i\pi$.
Based on the results obtained above, we can get the critical acceleration for chiral symmetry restoration $a_c(\Omega)$ as a function of angular velocity $\Omega$ analytically, by setting $m \to 0$.
Following the procedure outlined, the expression for $F_\O$ in Eq.~(\ref{eq:ft1}) consequently takes the form
\begin{equation}
\begin{aligned}
        F_\Omega&=i\frac{1}{\pi^3} \sum_{n=1} (-1)^n \int_{C} \ud\lambda \left( \frac{1}{i a \beta  n+\lambda  } \right) \cosh(\frac{\l}{2}) \frac{1}{{(2-2\cosh(\l))/a^2+r^2(2-2\cos(n\b\O_I))}} \cos(\frac{n\b \O_I}{2}) \\
        &=\sum_{n=1} (-1)^n\frac{2 a \cot \left(\frac{\beta  n \Omega_I }{2}\right) \sinh ^{-1}\left(a r \sin \left(\frac{\beta  n \Omega_I }{2}\right)\right)}{\pi ^2 a^2 \beta ^2 n^2 r+4 \pi ^2 r \sinh ^{-1}\left(ar\sin \left(\frac{\beta  n \Omega_I }{2}\right)\right)^2},
\end{aligned}
\end{equation}
where the $\lambda$ integration is analytically computed through the residue theorem. To perform the $n$ summation, we apply the fermionic analytic distillation theorem in \cite{Palermo:2021hlf}, yielding
\begin{equation}
    F_\Omega=-\left(\frac{\O_I ^2 \left(a^2 r^4 \O_I ^2-a^2 r^2-r^2 \O_I ^2-3\right)}{24 \pi ^2 (r^2 \O_I^2 +1)^2}+\frac{1}{6 \beta ^2 (r^2 \O_I^2 +1) }\right).
\end{equation}
By implementing the condition $\beta=2\pi/a$ and $\Omega_I \to -i \Omega$, $F_\O$ is finally simplified into
\begin{equation}
    F_\Omega=-\frac{a^2 \left(r^4 \Omega ^4+1\right)-r^2 \Omega ^4+3 \Omega ^2}{24 \pi ^2 \left(r^2 \Omega ^2-1\right)^2}.
\end{equation}
This directly leads to the gap equation
\begin{equation}
    G_\pi\left(\frac{\Lambda^2}{2\pi^2}-\frac{a^2 \left(r^4 \Omega ^4+1\right)-r^2 \Omega ^4+3 \Omega ^2}{24 \pi ^2 \left(r^2 \Omega ^2-1\right)^2}\right)=1,
\end{equation}
as given by Eq.~(\ref{eq:analyticalgap}) in the main text.

\bibliography{ref}

\end{document}